\documentclass[usenatbib]{mnras}

\usepackage{newtxtext,newtxmath}
\usepackage[T1]{fontenc}
\usepackage{ae,aecompl}

\usepackage{graphicx}	% Including figure files
\usepackage{amsmath}	% Advanced maths commands
\usepackage[final]{pdfpages}

\def\oiii{[O\,{\sc iii}]}

\def\h2{H$_2$}
\def\p1{Paper~I}

\def\kms{$\rm km\,s^{-1}$}
\def\str{{\sc starlight}}
\def\mc{$\mu$m} 
\title[Stellar population of nearby AGNs]{Gemini NIFS survey of feeding and feedback processes in nearby Active Galaxies: VI - Stellar Populations}

\author[Riffel, R. et al.]{Rog\'erio Riffel$^{1}$\thanks{E-mail: riffel@ufrgs.br}, 
Luis G. Dahmer-Hahn$^{2,10}$, 
Rogemar A. Riffel$^{3}$,
Thaisa Storchi-Bergmann$^{1}$, 
\newauthor
Natacha Z. Dametto$^4$, 
Richard Davies$^5$,
Leonard Burtscher$^6$,
Marina Bianchin$^{3}$,
\newauthor
Daniel Ruschel-Dutra$^7$,
Claudio Ricci$^{8,9}$
David J. Rosario$^{11}$\\
$^{1}$ Departamento de Astronomia, Instituto de F\'\i sica, Universidade Federal do Rio Grande do Sul, CP 15051, 91501-970, Porto Alegre, RS, Brazil \\
$^2$Shanghai Astronomical Observatory, Chinese Academy of Sciences, 80 Nandan Road, Shanghai 200030, China\\$^{3}$ Departamento de F\'\i sica, Centro de Ci\^encias Naturais e Exatas, Universidade Federal de Santa Maria, 97105-900, Santa Maria, RS, Brazil \\ 
$^{4}$ Centro de Astronomia (CITEVA), Universidad de Antofagasta, Avenida Angamos 601, Antofagasta, Chile\\
$^5$Max-Planck-Institut f\"ur Extraterrestrische Physik, Postfach 1312, 85741, Garching, Germany\\
$^6$Leiden Observatory, PO Box 9513, 2300 RA, Leiden, The Netherlands\\
$^7$Departamento de F\'isica, Universidade Federal de Santa Catarina, P.O. Box 476, 88040-900, Florian\'opolis, SC, Brazil\\
$^8$ N\'ucleo de Astronom\'ia de la Facultad de Ingenier\'ia, Universidad Diego Portales, Av. Ej\'ercito Libertador 441, Santiago, Chile\\
$^9$ Kavli Institute for Astronomy and Astrophysics, Peking University, Beijing 100871, China\\
$^{10}$Laborat\'orio Nacional de Astrof\'isica/MCTIC, Rua dos Estados Unidos, 154, Bairro das Na\c{c}\~oes, Itajub\'a, MG, Brazil\\
$^{11}$ School of Mathematics, Statistics \& Physics, Newcastle University, Newcastle upon Tyne, NE1 7RU, UK\\
}

\date{Accepted XXX. Received YYY; in original form ZZZ}

\pubyear{}

\hypersetup{draft}
\begin{document}
\label{firstpage}
\pagerange{\pageref{firstpage}--\pageref{lastpage}}
\maketitle

\label{firstpage}

\begin{abstract}

We use Gemini Near-Infrared Integral Field Spectrograph (NIFS) adaptive optics assisted data-cubes to map the stellar population of the inner few hundred parsecs of a sample of 18 nearby Seyfert galaxies. 
The near infrared light is dominated by the contribution of young to intermediate old stellar populations, with light-weighted mean ages $<t>_L\  \lesssim $ 1.5\,Gyr. Hot dust ($HD$) emission is centrally peaked (in the unresolved nucleus), but it is also needed to reproduce the continuum beyond the nucleus in nearly half of the sample. We have analysed the stellar population properties of the nuclear region and their relation with more global properties of the galaxies.  We find a correlation between the X-ray luminosity and the contributions from the $HD$, featureless continuum $FC$ and reddening $A_V$. We attribute these correlations to the fact that all these properties are linked to the mass accretion rate to the active galactic nuclei (AGN).  We also find a correlation of the bolometric luminosity $log(LBol_{obs})$ with the mass-weighted mean age of the stellar population, interpreted as due  a delay between the formation of new stars and the triggering/feeding of the AGN. The gas reaching the supermassive black hole is probably originated 
from mass loss from the already somewhat evolved intermediate age stellar population ($<t>_L \lesssim $ 1.5\,Gyr). In summary, our results show that there is a significant fraction of young to intermediate age stellar populations in the inner few 100\,pc of active galaxies, suggesting that this region is facing a rejuvenation process in which the AGN, once triggered, precludes further star formation, in the sense that it  can be associated with the lack of new star formation in the nuclear region.

\end{abstract}

%________________________________________________________________

\section{Introduction}

%\rrnotes{Need to be polished}

Since the pioneering studies linking the mass of supermassive black holes (SMBH) with the velocity dispersion of their host galaxies bulges it has become accepted that the products of active galactic nuclei (AGN) accretion and star-formation
are related \citep[e.g.][]{Magorrian+98,Ferrarese+00,Gebhardt+00}. It is also accepted that nuclear star formation and  
 AGN can coexist in the inner region of galaxies, suggesting that the growth of SMBH (by gas accretion) and galaxies (by star formation) are coupled \citep[][for a review]{Heckman+14,Madau+14}. 
In the modern view of galaxy evolution, it is well established that AGN feedback plays a fundamental role by impacting (quenching, suppressing or triggering) star formation from nuclear to galactic scales \citep[e.g.][]{DiMatteo+05,Hopkins+10a,Harrison+17,Storchi-Bergmann+19,Riffel+21,Ellison+21}. 

Theoretical studies and simulations of gas inflows around galactic nuclei lead to episodes of star formation in the nuclear region  \citep[$r\lesssim$ 100~pc,][ for example]{Kormendy+13a}. In this sense, one of the  most popular processes invoked to regulate star formation is the AGN feedback \citep[][and references therein]{Terrazas+20}. The accretion of matter into a SMBH, that leads to the triggering of an AGN, can in fact inject enough
energy into the galaxy affecting its star formation by heating (or removing) the gas \citep[e.g.][]{Granato+04,Zubovas+13,Zubovas+17a,Fabian+12,King+15,Trussler+20} or, as suggested in other studies, 
 these outflows and jets, can in some cases (depending on the AGN luminosity), compress the galactic gas, and therefore act as a catalyzer, boosting the star formation \citep[e.g.][]{Rees+89,Hopkins+12,Nayakshin+12,Bieri+16,Zubovas+13,Zubovas+17a} and even form stars inside the outflow \citep[e.g.][for observational examples, see \citet{Maiolino+17,Gallagher+19}]{Ishibashi+12,Zubovas+13,El-Badry+16,Wang+18}.
 
 It is also well established that cosmological simulations performed without the inclusion of feedback effects are not able to reproduce the low and high luminosity ends of the galaxy luminosity function  \citep[e.g.][]{Springel+05,Vogelsberger+14,Crain+15}, and underestimate the ages of the stars of the most massive galaxies when compared with observations \citep[][]{Croton+06}. These results clearly show 
that an effective feedback is required to reproduce the galaxy properties, but simulations can only provide limited insight on the nature and source of the feedback processes \citep[AGN or supernova dominated, ][]{Schaye+15}. This is because the observations are limited and do not provide strong enough constraints to be used in the modelling. In particular, in the case of AGN hosts it is necessary to robustly quantify the star formation history in the inner region of AGN hosts to see if star formation is boosted or quenched  \citep[][]{Riffel+21}.

Observations have clearly shown that nuclear star formation is common in AGNs \citep[e.g.][]{Terlevich+90a,Storchi-Bergmann+01a,CidFernandes+04,Riffel+07,Riffel+09,Storchi-Bergmann+12a,Esquej+14a,Riffel+16a,Ruschel-Dutra+17a,Hennig+18,Mallmann+18,Riffel+21,Burtscher+21}. Since AGN activity is most likely intermittent on time scales $\sim$ 0.1 - 1\,Myr \citep{Novak+11,Schawinski+15}, while the timescale of star formation in the host galaxy is much longer \citep[$\sim$100\,Myr, ][]{Hickox+14,Burtscher+21}, it is very difficult to compare the AGN life-cycle with nuclear starbursts ages (which have an uncertainty that may be larger than the time to trigger AGN activity). Therefore, these types of studies are rare and their results inconclusive. 

 The triggering mechanism of nuclear activity is still unclear \citep[see][for a review]{Alexander+12,Storchi-Bergmann+19}. Over the years many mechanisms were discussed as being the drivers (or not) of AGN activity, for example: galactic environment \citep[e.g.][]{Mauduit+07,Marshall+18,Davies+17}, mergers \citep[e.g.][and references therein]{Villforth+14,Goulding+18,Pan+19,Gao+20,Marian+20}, bar instabilities \citep[e.g.][]{Knapen+00}, accretion of gas clouds \citep[e.g.][]{Maccagni+14} and dust lanes \citep[][]{Lopes+07,Prieto+19}.
 
 In addition, a dusty torus, rich in molecular gas, is located in the inner parts of the host galaxy  \citep[e.g.][]{Antonucci+93,Rodriguez-Ardila+04,Rodriguez-Ardila+05,Riffel+13,Riffel+21a,Ramos-Almeida+17,BewketuBelete+21,Garcia-Burillo+19,Garcia-Burillo+21}. Such a huge gas reservoir could be responsible for nuclear activity \citep[][]{Reichard+09},
 and since the most important ingredient for star formation is the availability of a large amount of cold gas \citep[][]{Kennicutt+12}, the same gas fuelling the AGN could be used to form stars.  However, there is no consensus on whether AGN fuelling occurs at the same time as star formation \citep{Kawakatu+08a}, follows it during a post-starburst phase \citep{Davies+07a,Riffel+09} or there is a lack of any association with recent star formation \citep{Sarzi+07a}.

 Literature results trying to connect the AGN and star-formation are controversial. Studies suggest that the fraction of young stars in the vicinity of AGN is usually related with the luminosity of the AGN, with the most luminous AGNs presenting larger fractions of young stellar populations \citep[e.g.][]{Riffel+09,Ellison+16,Ruschel-Dutra+17a,Zubovas+17a, Mallmann+18}. Less efficient processes may be sufficient to supply lower accretion rates. \citet{Davies+07a}, for example, has shown that the age of the starburst is associated with the AGN luminosity. On the other hand in \citet{Burtscher+21}  no relation of the X-ray luminosity with the fraction of young stellar population was found. Indeed, many of the confusing or inconclusive results about local AGN may be attributable to the luminosity being a `third parameter': previous analyses of local AGN have typically been performed for objects with low luminosities \citep[e.g.][]{Davies+15a}.

It is, then, fundamental to this debate to investigate if there is an association between star formation and nuclear activity, by mapping the stellar population properties in the inner few tens of parsec around AGNs over a wide range of AGN luminosities and in a wavelength region sensitive to the AGN featureless components (accretion disc and hot dust emission) and the stellar population at the same time. The ideal spectral region for this is the near-infrared (NIR = $\sim$0.8-2.4\mc) where the AGN featureless continuum (FC), the hot dust (HD) and the stellar population components can be fitted together and disentangled \citep{Riffel+09,Riffel+11c}.

Here we map the stellar population components in the inner few hundred parsecs of a volume limited, X-ray luminosity selected and adaptive optics (AO) assisted observed sample of AGNs, the {\it Gemini NIFS survey of feeding and feedback processes in nearby Active Galaxies} \citep[][]{Schonell+17,Riffel+17,Riffel+18,Riffel+21a, Bianchin+22}. This paper is structured as follows: in \S~\ref{obs} we present the sample, observations, and data reduction. The stellar population synthesis procedures are presented in \S~\ref{popstar}. Results  are presented in \S~\ref{results}. The discussion is made in \S~\ref{discussion} while the conclusions are presented in \S~\ref{conc}. We have adopted $H_0$ = 69.32 $\rm km s^{-1} Mpc^{-1}$ from the Nine-year Wilkinson Microwave Anisotropy Probe (WMAP) observations \citet{Hinshaw+13} available in {\sc astropy} package \citep{AstropyCollaboration+18}.

\section{Sample and observations}\label{obs}

\subsection{The sample}

The present sample is composed of the galaxies in the Gemini NIFS survey of feeding and feedback processes in nearby active galaxies \citep{Riffel+17,Riffel+18} that have observations in the $J$ and $K$ (or $K_s$) bands. In short, the sample was selected using the  Swift-BAT 60-month catalogue \citep{Ajello+12}, and  selected nearby galaxies with 14--195\,keV luminosities $L_X \ge 10^{41.5}$ ergs\,s$^{-1}$. Such selection guarantees a high completeness, since hard X-ray selected samples are not significantly biased against obscuration, up to column density of $N_{\rm H}\simeq 10^{24}\rm\,cm^{-2}$ \citep[see Fig.~1 of][]{Ricci+15}.
 As additional criteria, the object must be accessible for Gemini NIFS ($-30^\circ< \delta < 73^\circ$), and either its nucleus must be bright/pointy enough to guide the adaptive optics system during the observations or there must be a natural guide star available in the field for the same purpose. Finally, we only have included in the sample galaxies already previously observed in the optical and with extended \oiii$\lambda5007$ emission  available in the literature. This sample was completed with data already observed by the AGNIFS team in other projects. For additional details see \citet{Riffel+18}. Our final sample is composed of 18 Local Universe (0.00303 $\leq z\leq $0.02213; 13.10\,Mpc $\leq d\leq$ 95.71\,Mpc)  sources, listed in Tab.~\ref{tab_obs}. It includes six galaxies with results on the stellar populations already published by the AGNIFS team \citep{Riffel+10,Riffel+11,Storchi-Bergmann+12a,Schonell+17,Dahmer-Hahn+19a,Dahmer-Hahn+19,Diniz+19}. These galaxies are identified  with the $\dagger$ symbol after their names. 

In order to compare the stellar population properties with more general ones, we have also collected or computed additional properties for the galaxies, as follows:
\begin{enumerate}
   
\item  Observed X-ray luminosity (2-10 keV) were taken from the BASS survey DR1 \citep{Koss+17,Ricci+17}. The exceptions are: Mrk\,1157,  Mrk\,1066
and  NGC\,5929  that are from \citet{Cardamone+07} and Mrk\,607 from \citet{LaMassa+11}. 

\item Intrinsic X-ray luminosities were also collected from the BASS survey DR1, except for Mrk\,1157, Mrk\,1066,  NGC\,5929 and Mrk\,607  where we have calculated their values as follows. For all sources we collected archival data, which includes {\it XMM-Newton} (Mrk\,607,  Mrk\,1066 and Mrk\,1157), {\it Chandra} (NGC 5929), {\it Swift/BAT} (Mrk\,1066) and {\it NuSTAR} (Mrk\,607 and NGC\,5929). The spectra were reduced and fitted following the same approach outlined in \citet{Ricci+17}. We found that all objects are heavily obscured, with three of them (Mrk\,607, Mrk\,1066 and Mrk\,1157) having Compton-thick ($N_{\rm H}\geq 10^{24}\rm\,cm^{-2}$) column densities.

\item The bolometric luminosity was calculated using the observed   X-ray luminosity together with the equation  \citep{Ichikawa+17}:
\begin{equation*}
  \log L_{\rm bol} = 0.0378 (\log L_{2-10})^2 - 2.00 \log L_{2-10} +60.5.
  \end{equation*}
\item Stellar velocity dispersion ($\sigma$) was taken from \citet{Riffel+17} except for: NGC\,1068 \citep{Storchi-Bergmann+12a}, NGC\,4151
 \citep{Ho+09}, NGC\,1125 \citep{Garcia-Rissmann+05}, NGC\,5506 \citep{Oliva+99} and Mrk\,79 \citep{Greene+06}.

\item The supermassive black hole mass was computed using the $M-\sigma$ relation from \citet{Caglar+20}:
\begin{equation*}
\log\left( \frac{M_\bullet}{M_\odot} \right) = (8.14\pm0.20) + (3.38\pm0.65) \log \left(\frac{\sigma_e}{ \mathrm{200\,km\,s}^{-1}} \right).
\end{equation*}

\item The Eddington luminosity \citep{Rybicki+79}: 
\begin{equation*}
L_{\rm Edd}=1.26\cdot 10^{46}\,\text{erg s}^{-1} \frac{M_\bullet}{10^8 M_\odot}.
\end{equation*} 

\item Eddington ratio was obtained from $L_{Bol}$ and calculated using the observed $L_X$ ($\lambda_{obs} = L_{Bol}^{obs}/L_{\rm Edd}$) and intrinsic X-ray luminosity ($\lambda_{int} = L_{Bol}^{int}/L_{\rm Edd}$)
\item  $\rm H_2$ and H\,{\sc ii} masses in the inner 125\,pc were taken from \citet{Riffel+21a}.

\end{enumerate}

\begin{table*}
\renewcommand{\tabcolsep}{0.70mm}
\centering
\small
\caption{Sample properties properties }
\vspace{0.3cm}
\begin{tabular}{l c c l l c c c c c c c c c c c c }
\hline
\hline
Galaxy  & z & d & Nuc. Act. & Hub. Type & $L_{X}^{obs}$ & $L_X^{int}$ & $L_{Bol}^{obs}$& $L_{Bol}^{int}$ & $\sigma_{\star}$    & $M_{\bullet}$  &$L_{Edd}$&    $\lambda_{obs}$ & $\lambda_{int}$   &   $M(H_2)$          &     M(H{\sc ii})    \\ 
(1)     &    (2)       &     (3)      &  (4)        &   (5)                     &  (6)        &   (7)                  & (8)     &   (9)              &    (10)           &  (11)            &   (12)  &   (13)   &  (14)   &   (15) & (16) \\
\noalign{\smallskip}
\hline	
NGC\,788              &  0.01365  &   59.03  & Sy2   & SA0/a?(s)	  &  42.13  	&	43.02		&	43.33		& 44.42 	  & 187$\pm$4	&	  8.04$\pm$0.07   & 46.14	 &    1.55E-03     &   1.91E-02 	 &  12.81$\pm$0.46  &  19.54$\pm$0.6		\\
NGC\,1052$\dagger$    &  0.00446  &   19.29  & Sy2   & E4			  &   41.36 	&	 41.52  	&	 42.45  	&  42.63	  & 245$\pm$4	&	  8.44$\pm$0.06   & 46.54	 &    8.13E-05     &   1.23E-04 	 &  11.02$\pm$1.14  &  5.67$\pm$1.63	   \\  
Mrk\,79               &  0.02213  &   95.71  & Sy1.5 & SBb  		  &  42.93  	&	43.10		&	44.30		& 44.52 	  & 130$\pm$10  &	  7.51$\pm$0.21   & 45.61	 &    4.90E-02     &   8.13E-02 	 &  52.9$\pm$2.09	& 23.86$\pm$2.49	 \\
NGC\,1125             &  0.01128  &   48.78  & Sy2   & (R')SB(r)0/a?&   41.06 	&	 42.76  	&	 42.10  	&  44.10	  & 118$\pm$9	&	  7.37$\pm$0.21   & 45.47	 &    4.27E-04     &   4.27E-02 	 &  21.26$\pm$0.69  &  64.11$\pm$0.83		\\  
NGC\,1068$\dagger$    &  0.00303  &   13.10  &  Sy2  & (R)SA(rs)b	  &   41.00 	&	 42.74  	&	 42.05  	&  44.07	  & 162$\pm$5	&	  7.83$\pm$0.09   & 45.93	 &    1.32E-04     &   1.38E-02 	 &   202.28$\pm$2.19&  191.95$\pm$3.84   \\  
NGC\,2110$\dagger$    &  0.00739  &   31.96  & Sy2   & SAB0$^-$ 	  &   42.48 	&	 42.65  	&	 43.75  	&  43.96	  & 238$\pm$5	&	  8.40$\pm$0.07   & 46.50	 &    1.78E-03     &   2.88E-03 	 &  41.71$\pm$0.84  &  16.75$\pm$0.98	  \\  
NGC\,3227             &  0.00329  &   14.23  & Sy1.5 & SAB(s)a pec    &   41.95 	&	 41.95  	&	 43.12  	&  43.13	  & 130$\pm$7	&	  7.51$\pm$0.14   & 45.61	 &    3.24E-03     &   3.31E-03 	 &  52.11$\pm$1.3	&  11.52$\pm$1.23	\\  
NGC\,3516             &  0.00871  &   37.67  & Sy1.5 & (R)SB0$^0$?(s) &   42.67 	&	 42.72  	&	 43.98  	&  44.04	  & 186$\pm$3	&	  8.03$\pm$0.06   & 46.13	 &    7.08E-03     &   8.13E-03 	 &  8.69$\pm$2.19	&  1.97$\pm$0.34	 \\  
NGC\,4151             &  0.00314  &   13.58  & Sy1   & SAB(rs)bc	  &   42.02 	&	 42.27  	&	 43.20  	&  43.50	  & 97$\pm$3	&	  7.08$\pm$0.07   & 45.18	 &    1.05E-02     &   2.09E-02 	 &   36.89$\pm$1.31 &  54.77$\pm$1.54	\\  
NGC\,4235             &  0.00757  &   32.74  & Sy1   & SA(s)a edge-on &   41.56 	&	 41.56  	&	 42.66  	&  42.66	  & 183$\pm$12  &	  8.01$\pm$0.20   & 46.11	 &    3.55E-04     &   3.55E-04 	 &  3.98$\pm$0.5	&  0.19$\pm$0.07	 \\  
NGC\,5506             &  0.00609  &   26.34  & Sy1.9 & Sa pec edge-on &   42.89 	&	 42.98  	&	 44.25  	&  44.37	  & 180$\pm$20  &	  7.99$\pm$0.37   & 46.09	 &    1.45E-02     &   1.91E-02 	 &  45.71$\pm$0.72  &  174.09$\pm$1.41   \\  
NGC\,5548$\dagger$    &  0.01673  &   72.35  & Sy1   & (R')SA0/a(s)   &   43.08 	&	 43.12  	&	 44.49  	&  44.54	  & 276$\pm$22  &	  8.61$\pm$0.28   & 46.71	 &    6.03E-03     &   6.76E-03 	 &   33.29$\pm$5.54 &  32.06$\pm$6.76	\\   
NGC\,5899             &  0.00844  &   36.50  & Sy2   & SAB(rs)c 	  &   41.91 	&	 42.21  	&	 43.07  	&  43.42	  & 147$\pm$9	&	  7.69$\pm$0.17   & 45.79	 &    1.91E-03     &   4.27E-03 	 &  17.79$\pm$0.8	&  6.58$\pm$1.09	 \\  
NGC\,5929             &  0.00831  &   35.94  & Sy2   & Sab? pec 	  &   40.10 	&	 41.58  	&	 41.08  	&  42.69	  & 134$\pm$5	&	  7.55$\pm$0.10   & 45.65	 &    2.69E-05     &   1.10E-03 	 &  17.1$\pm$0.9 	&  11.4$\pm$0.5				 \\
Mrk\,607              &  0.00888  &   38.40  & Sy2   & Sa? edge-on    &  41.83  	&	43.73		&	42.98		& 45.32 	  & 132$\pm$4	&	  7.53$\pm$0.08   & 45.63	 &    2.24E-03     &   4.90E-01 	 &  7.1$\pm$0.41	&  11.3$\pm$0.74	  \\
Mrk\,766              &  0.01292  &   55.88  & Sy1.5 & (R')SB(s)a?    &  42.69  	&	42.72		&	44.01		& 44.04 	  & 78$\pm$6	&	  6.76$\pm$0.18   & 44.86	 &    1.41E-01     &   1.51E-01 	 &  23.28$\pm$1.95  &  112.21$\pm$3.69    \\
Mrk\,1066$\dagger$    &  0.01202  &   51.98  & Sy2   & (R)SB0$^+$(s)  &   40.72 	&	 42.63  	&	 41.74  	&  43.94	  & 103$\pm$4	&	  7.17$\pm$0.09   & 45.27	 &    2.95E-04     &   4.68E-02 	 &   57.33$\pm$0.44 &  68.86$\pm$0.73	\\  
Mrk\,1157$\dagger$    &  0.01517  &   65.61  & Sy2   &(R')SB0/a 	  &    40.29	&	  42.87 	&	  41.28 	&	44.22	  & 92$\pm$4	&	  7.00$\pm$0.10   & 45.10	 &    1.51E-04     &   1.32E-01 	 & 9.80$\pm$1.9 	& 53.3$\pm$1.3		 \\
\noalign{\smallskip}
\hline
\end{tabular}
\label{tab_obs}
%\end{scriptsize}
\begin{list}{Table Notes:}
\item $\dagger$ Stellar population studies previously published by the ANGIFS team.
(1) Galaxy's name; 
(2) Distance in Mpc;
(3) Redshift taken from the BASS survey DR1 \citep[][http://www.bass-survey.com/]{Ricci+17,Koss+17,Oh+18}; 
(4) Nuclear activity;
(5) Hubble  type as quoted in NED\footnote{The NASA/IPAC Extragalactic Database (NED) is funded by the National Aeronautics and Space Administration and operated by the California Institute of Technology.};
(6) Logarithm of the observed hard X-ray luminosity (14-195 keV, in erg s$^{-1}$) from the BASS survey DR1, except for Mrk\,1157,  Mrk\,1066
and  NGC\,5929  that are from \citet{Cardamone+07} and Mrk\,607 that are from \citet{LaMassa+11};
(7)  Logarithm of the intrinsic hard X-ray luminosity (14-195 keV, in erg s$^{-1}$) from the BASS survey DR1, except for Mrk\,1157, 
Mrk\,1066,  NGC\,5929 and Mrk\,607  where we have calculated their values;
(8)  Logarithm of the bolometric luminosity calculated using the observed  hard X-ray luminosity (in erg s$^{-1}$);
(9)  Logarithm of the bolometric luminosity calculated using the intrinsic  hard X-ray luminosity (in erg s$^{-1}$);
(10) $\sigma \rm (kms{-1})$ taken from \citet{Riffel+17} except for: NGC\,1068 \citep{Storchi-Bergmann+12a}, NGC\,4151
\citep{Ho+09}, NGC\,1125 \citep{Garcia-Rissmann+05}, NGC\,5506 \citep{Oliva+99} and Mrk\,79 \citep{Greene+06};
(11)  Logarithm of the supermassive black hole mass in M$\odot$ computed using the $M-\sigma$ relation from \citet{Caglar+20};
(12)  Logarithm of the Eddington luminosity \citep{Rybicki+79};
(13) Edditngton ratio computed with  $L_{Bol}^{obs}$;
(14) Edditngton ratio computed with $ L_{Bol}^{int}$;
(15) $\rm H_2$ mass (in $\rm 10^{1} M\odot$) in the inner 125\,pc taken from \citet{Riffel+21}.
(16) H\,{\sc ii} mass (in $\rm 10^{4}M\odot$) in the inner 125\,pc taken from \citet{Riffel+21}.
\end{list}
\end{table*}

\subsection{Observations}

The observations of our sample were done with the Gemini Near-Infrared Integral Field Spectrograph \citep[NIFS,][]{McGregor+03}. All observations were performed using the Gemini North Adaptive Optics system ALTAIR from 2006 to 2018, except for NGC\,1125 that was observed in the seeing limited mode \citep{Riffel+18,Riffel+21a}.  The observations followed an Object-Sky-Object dither sequence, with sky observations made off-source since all  targets are extended. Individual exposure times varied according to the target. The data comprise J and K (K$_l$)-band observations at angular resolutions in the range 0\farcs12--0\farcs20, depending on the performance of the adaptive optics module. Additional details can be found in \citet{Riffel+18}.

The angular resolution for each galaxy was derived from the $FWHM$ of the flux distribution of the standard stars, corresponding to few tens of parsecs at the distance of most galaxies as presented in \citet{Riffel+17} and \citet{Riffel+21a}.

\subsection{Data reduction}

 The data reduction followed the standard procedure and was performed using tasks that were specifically developed for NIFS data reduction, contained in the {\sc nifs} package, which is part of the {\sc gemini iraf} package, as well as generic {\sc iraf} tasks \citep{Tody+86,Tody+93}. The procedures included the trimming of the images, flat-fielding, sky subtraction, wavelength and s-distortion calibrations. 
 
 The telluric absorption corrections have been performed using the spectra of telluric A-type standard stars, observed just before/after the observation of each galaxy.  They were also used to flux calibrate the spectra of the galaxies by interpolating a black body function to their spectra. Finally, datacubes were created for each individual exposure at an angular sampling of 0\farcs15$\times$0\farcs15 and combined in a final datacube for each galaxy. All datacubes cover the inner $\sim3^{\prime\prime}\times3^{\prime\prime}$, with exception of the datacube for NGC\,4151 that covers the inner $\sim3^{\prime\prime}\times4^{\prime\prime}$,  due to spatial dithering during the observations \citep{Storchi-Bergmann+09a, Storchi-Bergmann+10a} and for NGC\,1068, covering the inner $5^{\prime\prime}\times5^{\prime\prime}$ \citep{Storchi-Bergmann+12a,Riffel+14,Barbosa+14}

We followed the steps used in previous papers of the AGNIFS group \citep[e.g.][]{Riffel+11,Storchi-Bergmann+12a,Diniz+19,Dametto+19a,Dahmer-Hahn+19a},
 since the different bands of our data were acquired during different nights. We have cross calibrated the data-cubes using cross dispersed (XD) long-slit data presented by \citet{Riffel+06,Riffel+13a,Riffel+15,Riffel+19} and \citet{Mason+15a} or unpublished XD spectra (Riffel et al., in preparation).

\section{Stellar Population Synthesis}\label{popstar}

The galaxies' integrated spectra are composed of several components, such as the underlying stellar (and gas) continuum and  dust emission. In the case of active galaxies, besides the above, AGN torus and accretion disk components also need to be considered \citep{Riffel+09,Burtscher+15}. The stellar population synthesis consists of disentangling the percent contribution of these components from the integrated spectrum.  

It is worth mentioning that since there is a large number of parameters (e.g. age, metallicity, kinematics and reddening, and AGN components) to be fitted, many techniques have been employed over the years to disentangle the components of a galaxy spectrum \citep[for a review see][]{Walcher+11,Conroy+13}. Therefore, many fitting codes have been developed by different groups \citep[e.g.][]{CidFernandes+05,Ocvirk+06,Koleva+09,Tojeiro+07,Cappellari+17,Sanchez+16,Wilkinson+17,Gomes+17,Johnson+21} with their own priorities in mind \citep[see][for example]{Gomes+17}. Naturally, comparisons among the different codes have also been presented, and they show that in general the codes do produce consistent results when fitting the same data with the same ingredients \citep[e.g.][]{Koleva+08,Dias+10,Gomes+17,Goddard+17,Ge+18, CidFernandes+18}.

To perform the stellar population synthesis we applied {\sc starlight}\footnote{This choice was made mainly to produce self consistent results with the previous studies done by our team, allowing for easier comparisons of the results.} \citep{CidFernandes+04,CidFernandes+05,Asari+07,CidFernandes+18}. The code fits the full underlying absorption and continuum in the observed spectra as a combination of different proportions of the base of elements, excluding emission lines and spurious data, mixing computational techniques originally developed for semi-empirical population synthesis with ingredients from evolutionary synthesis models 
\citep{CidFernandes+04,CidFernandes+05}. In summary,  the code fits an observed spectrum, $O_{\lambda}$, with a combination, in different proportions, of $N_{\star}$ simple stellar populations (SSPs). The visual extinction, $A_V$, is modelled by \str\ as due to foreground dust. 
In the fits we use the CCM \citep{Cardelli+89} extinction law. 
The modeled spectrum, $M_{\lambda}$, is obtained through the following equation:

\begin{equation}
M_{\lambda}=M_{\lambda 0}\sum_{j=1}^{N_{\star}}x_j\,b_{j,\lambda}\,r_{\lambda}  \otimes G(v_{\star},\sigma_{\star})
\label{streq}
\end{equation}
where $x_j$ is the population vector, $b_{j,\lambda}$ is the $j$th base element, $\,r_{\lambda}$ is the 
reddening factor of the $j$th component normalised at $\lambda_0$, the reddening term is represented by 
$r_{\lambda}=10^{-0.4(A_{\lambda}-A_{\lambda 0})}$, $M_{\lambda 0}$ is the
synthetic flux at the normalisation wavelength (we have used $\rm \lambda_{norm}=21\,100$\AA\ in the rest frame).  The convolution 
operator is $\otimes$ and $G(v_{\star},\sigma_{\star})$ is the Gaussian distribution used to model the line-of-sight velocity distributions of the stars, which is centred at velocity $v_{\star}$  with dispersion  $\sigma_{\star}$. 
The final fit is carried out through a chi-square minimisation, as follows: 

\begin{equation}
\chi^2 = \sum_{\lambda}[(O_{\lambda}-M_{\lambda})w_{\lambda}]^2
\end{equation}
where emission lines and spurious features are excluded from the fit by fixing $w_{\lambda}$=0  at the corresponding wavelengths.

\begin{figure}
    \centering
    \includegraphics[scale=0.6]{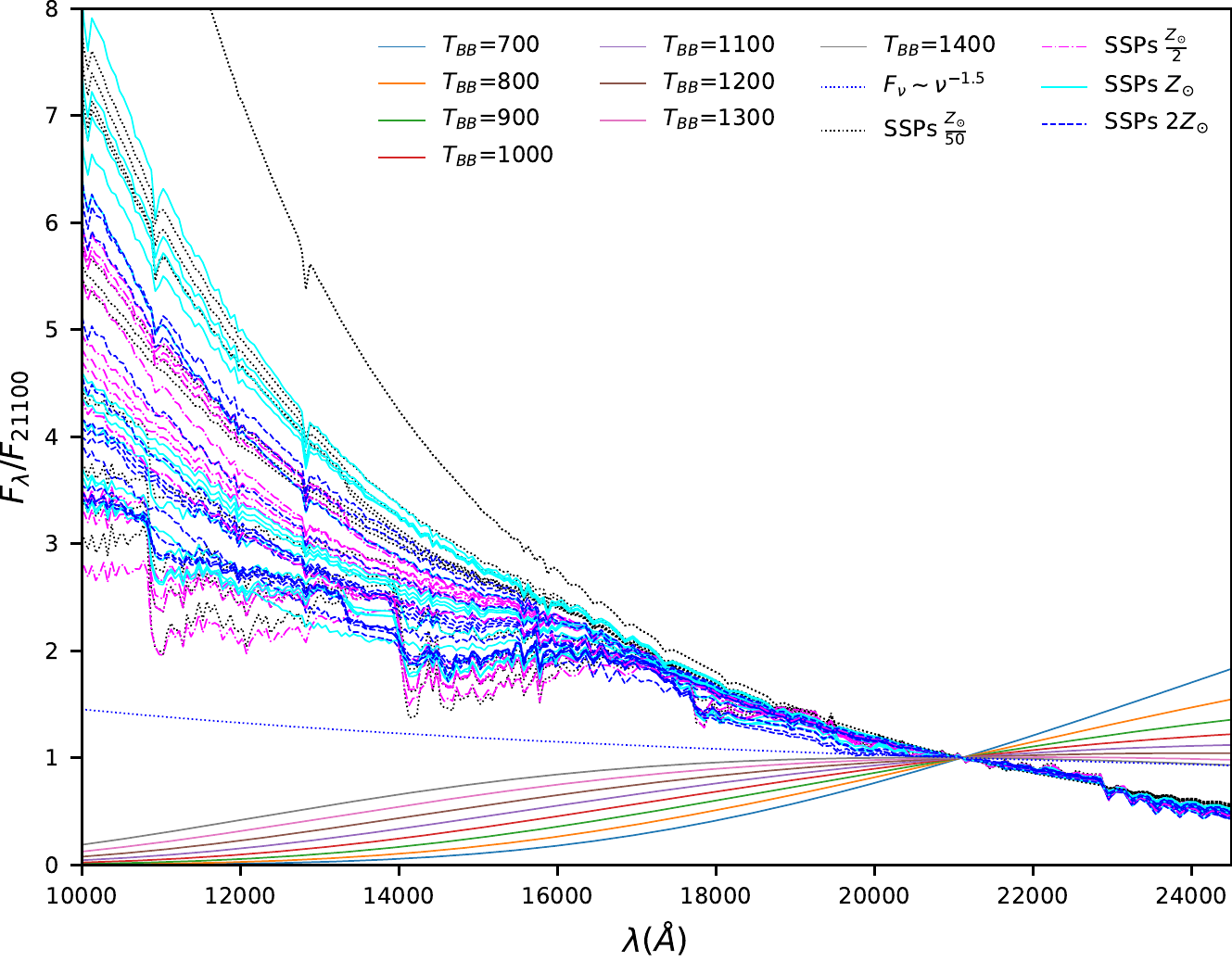}
    \caption{The base of elements comprises SSPs with 4 metalicities ($Z$ = 0.02, 0.5, 1 and 2 $Z_\odot$) and 12 ages (t = 0.01, 0.03, 0.05, 0.1, 0.2, 0.5, 0.7, 1, 2, 5, 9, 13~Gyr). The accretion disk (FC) is described as  a power-law of the form $F_\lambda \propto \lambda^{-0.5}$. Hot dust emission  component are described as eight Planck distributions
(black-body, BB), with temperature ranging from 700 to 1400 K, in steps of 100 K. The identification of the components are on the labels. For details see text.}
    \label{base}
\end{figure}

In the present paper we have followed \citet{Riffel+09,Riffel+11c} and used as simple stellar population (SSP) the evolutionary population synthesis (EPS) models of \citet{Maraston+05}. These models are able to foresee characteristic NIR absorption features \citep[e.g.][]{Riffel+07,Riffel+15} and to produce consistent results when compared with properties of the emission line gas in star forming regions of galaxies \citep{Dametto+19a}.  The base of elements comprises SSPs with 4 metalicities ($Z$ = 0.02, 0.5, 1 and 2 $Z_\odot$) and 12 ages (t = 0.01, 0.03, 0.05, 0.1, 0.2, 0.5, 0.7, 1, 2, 5, 9, 13~Gyr). To account for the accretion disk FC we have used a power-law of the form $F_\lambda \propto \lambda^{-0.5}$ \citep[e.g.][]{Koski+78,CidFernandes+05}. In order to properly account for the hot dust emission  component, eight Planck distributions
(black-body, BB), with temperature ranging from 700 to 1400 K, in steps of
100 K, were included in the fits. The lower limit was chosen because lower temperatues are hard to detect in this spectral range \citep{Riffel+09} and the upper limit is very close to the sublimation temperature of the dust grains \citep[e.g.][]{Barvainis+87,Granato+94}.  For more details on the effects of these components in the NIR spectra and on the definition of our base of elements see \citet{Riffel+09}. The components used in the fits are shown in Fig.~\ref{base}.

\section{Results} \label{results}

\begin{figure*}
    \centering
    \includegraphics[scale=0.6]{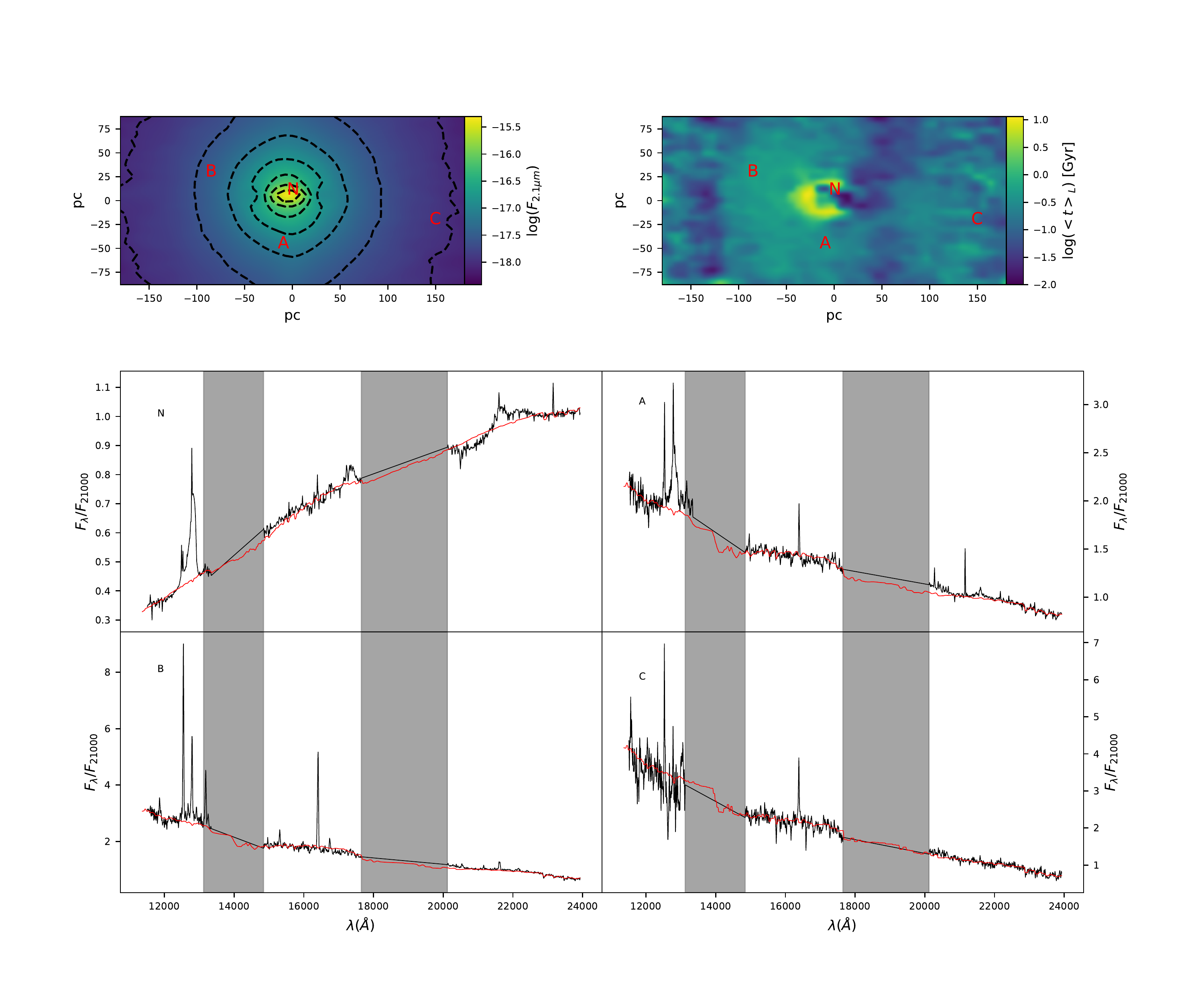}
    \caption{In top panels we show continuum (left) and  light-weighted mean age (right) maps for NCG~4151. In the bottom panels we show examples of four single spaxels fits: the nuclear region and other 3 distinct locations identified in the top panels. This is the only galaxy with J, H and K bands avaliable.}
    \label{fit_example}
\end{figure*}

Since \str\ is not designed to deal with data cubes, we have used the in-house {\sc megacube} tool \citep{Mallmann+18} as a front end for \str\ fits and to produce maps for the derived quantities. Therefore each single spaxel was corrected for galactic extinction using the \citet{Schlegel+98} dust maps and redshift corrected using the values listed in Tab.~\ref{tab_obs}. An example of individual spectrum fits can be seen in Fig~\ref{fit_example}.

To take into account the noise effects on the data that wash away differences between similar spectral components, as well as the age-metallicity degeneracy\footnote{The difference between two spectrally similar populations  may be in the noise level of the data; by binning them in a condensed population vector, we are left with a coarser but more robust fraction for each age bin. For a detailed test see \S3.2 of \citet{CidFernandes+04}.} \citep[e.g.][]{Worthey+94a}, we have followed \citet{CidFernandes+05,Riffel+09} and we have defined condensed population vectors and have produced maps and radial profiles for different properties, for each one of the galaxies, as follows: 
\begin{description}
    \item[$xy$:] sum of the percent  contribution of SSPs with ages $t\leq$ 50~Myr;
    
    \item[$xyo$:] sum of the percent contribution of SSPs with ages  in the range 100~Myr $\leq t \leq$ 200~Myr;
    
    \item[$xiy$:] sum of the percent contribution of SSPs with ages  in the range 500~Myr $\leq t \leq$ 700~Myr;
    
    \item[$xio$:] sum of the percent contribution SSPs with ages  in the range 1~Gyr $\leq t \leq$ 2~Gyr;
    
    \item[$xo$:] sum of the percent contribution 5~Gyr $\leq t \leq$ 13~Gyr;
    
    \item[HD:]  sum of the percent contribution of all Planck function components.
    
    \item[FC:] is the featureless  percent contribution.
    
    \item[$<t>_L$:] is the light weighted mean age, which is defined as follows \citep{CidFernandes+05}:
    \begin{equation}
    \langle {\rm log} t_{\star} \rangle_{L} = \displaystyle \sum^{N_{\star}}_{j=1} x_j {\rm log}t_j.
    \label{ewMage}
    \end{equation}
\end{description}
Note that all these values are in light fractions and are limited by the elements included in the base.

An example of the maps is shown in Fig.~\ref{maps0} (the remaining maps are in the supplementary material).  The red cross marks the peak of the continuum at the normalisation point. In the same plot, we show radial profiles for these properties.  The profiles shown in Fig.~\ref{maps0} were built using the {\sc megacube} tool \citep{Mallmann+18} and we used the standard 30 equally spaced (in azimuthal angle) radial profiles. These profiles, in the galaxy plane, have been limited to angular distances from the major axis of $\rm \theta_{max} = tan^{-1} (b/a)$ degrees, where a and b are, respectively, the mean values of the semi-major and semi-minor axis which were derived using the {\sc photutils.isophote Ellipse  astropy} affiliated package \citep{Bradley+19}. The solid line represents the mean value and the shaded region is its standard deviation. The dotted vertical line is the point spread function (PSF) of the full width at half maximum (FWHM) taken from \citet{Riffel+17,Riffel+21a}. To remove possible spurious data, not removed in the fits, we have recomputed the standard {\sc starlight}  $Adev_S$ value as follows: 

\begin{equation}
Adev_S = \frac{100}{N_l^{eff}} \sum \frac{| F_{obs} - F_{syn} |}{F_{syn}},
\end{equation}
where $N_l^{eff}$ is the number of points used in the fits, while, $F_{obs}$ and $F_{syn}$ are the observed and synthesised fluxes, respectively.  We have identified all points with $Adev_S >$10 (red lines) and $Adev_S >$20 (white lines). It is clear that almost all the spaxels in all analysed sources have $Adev_S \leq$10.   For the radial profiles, however, we used all spaxels, and the loose of quality on the fits is observed as a larger scatter around the mean values for large radii\footnote{The radial profiles have been obtained by averaging the individual (deprojected) profiles obtained along each one of the yellow lines (in both directions) in the top panel of the figures.}. 

For completeness, we have refitted the SPs of the six galaxies already studied by the AGNIFS team and we present the corresponding maps and profiles as supplementary material.

\subsection{Individual Results}

\begin{description}
   \item[{\bf NGC~788:}] The light of this galaxy shows an important contribution of the two younger ($xy$ and $xyo$) components with a maximum value of $\sim$ 40 percent. The radial profiles show that these two components only show up in the outer parts of the field of view (FoV). It is, however, worth mentioning that there is a ``stripe'' on the east side, that can be due to an instrumental fingerprint (not completely removed in the data reduction process), therefore this should just be taken as a trend. It is clear from these maps that the light of the central region of this galaxy is dominated by stars with ages  in the range 500~Myr $\leq t \leq$ 700~Myr ($xiy$) with contributions $>$ 40 percent over the FoV. A fraction of $\sim$25 percent of older stars ($xio$ and $xo$ stellar population component [SPC]) are detected throughout the whole FoV. Hot dust emission contributes 35 percent of the nuclear emission of this source, while no FC is required. The mean age shows a negative profile with the central region having older stellar populations than the outer parts. 
  
  \item[{\bf NGC~1052:}] This source is dominated by intermediate and old age stellar populations. A very tiny fraction of $FC$ is found in the unresolved nucleus. The stellar content of this galaxy is discussed in detail in \citet{Dahmer-Hahn+19a,Dahmer-Hahn+19}.

  \item[{\bf Mrk~79:}] A clear ring-like structure is observed in the $xy$ and $xyo$ SPC, with a fraction reaching 50 percent for the $xyo$ population. Negligible contributions of the intermediate age components are observed, with a smoother central distribution of older stars ($xo$) surrounded by the $xyo$ ring-like (r $\ge$ 300\,pc) SPC. Hot dust emission is observed over the entire FoV showing a maximum in the nuclear region. A FC emission of $\sim$40 percent is detected in the nuclear region. In the mean age map we see that the central region of this galaxy is dominated by a 2~Gyr population. 
  
  \item[{\bf NGC~1125:}] This galaxy presents a complex star formation history (SFH). Locations 100~pc from the centre towards the NW are dominated by a $xy$ SPC, reaching values up to 50 percent.  A ring-like structure is observed in the $xyo$ map starting at $\sim$ 100\,pc from the centre with a fraction of $\sim$ 25 percent. This ring structure surrounds a ring of intermediate age stars ($xiy$), which dominates the light in the central region of this galaxy. Fractions of $\sim$ 25 percent for the two older components are also observed, being located in regions from the centre towards the North. A vertical stripe can be seen in the East direction in this $xo$ component (and as already mentioned before for other galaxies, it can be due to a residual instrumental fingerprint). Hot dust emission is seen in the unresolved centre (r$\le$100\,pc) of the FoV reaching 20 percent, and no FC emission is detected. The mean age map shows that the central region of this galaxy is dominated by populations with ages smaller than 1\,Gyr, decreasing outwards.
  
     \item[{\bf NGC~1068:}] This source has a very complex SFH, with the unresolved nuclear region being dominated by the $HD$ component. Outside this region, large fractions of the $HD$ are also observed. Its stellar content is dominated by young and intermediate age populations. A detailed analysis of the stellar populations of this galaxy is made in \citet{Martins+10a} and \citet{Storchi-Bergmann+12a}.

   \item[{\bf NGC~2110:}] In terms of stellar content the central region of this galaxy is dominated by intermediate and old stellar populations, with a smaller fraction in the $xy$ component. It also shows a large fraction of hot dust emission in the unresolved nucleus. A detailed study of the stellar populations of this galaxy is presented in \citep{Diniz+19}.

  \item[{\bf NGC~3227:}] Very small fractions of the two younger SPC are detected. An off-centred (x=60\,pc,y=-20\,pc) ring-like structure of intermediate young ($xiy$) stars can be seen as well as a significant fraction of $xio$ SPC is detected over the entire FoV. This structure has a very good agreement with the off-centre, ring-like structure found in CO(2-1) Atacama Large Millimeter/submillimeter Array (ALMA) observations by \citep{Alonso-Herrero+19,Alonso-Herrero+20}, as well as with a low velocity dispersion stellar ring \citep{Barbosa+09} and in $Pa_\alpha$ emission \citep{Bianchin+22}, indicating that this gas reservoir is forming stars. We note that a vertical stripe is seen to the W side of the FoV and that this feature should be treated with caution.
  The old SPC is also detected over the entire FoV contributing 25 percent, except for the unresolved nucleus (r$\le$15\,pc), which is dominated by hot dust emission (75 percent) together with 10 percent of FC emission. The mean age map shows that NGC~3227 is dominated by stellar populations with ages lower than 2~Gyr.
  
  \item[{\bf NGC~3516:}] A partial ring of the $xy$ SPC is observed from West to East towards the South locations, starting at 30\,pc out to the end of our FoV (150\,pc). No $xyo$ population was required. A clear ring (100\,pc\,$<$r$<$\,150\,pc) of $xiy$ stars is detected surrounding a ring of the $xio$ component (with a smaller contribution of $xo$). This ring surrounds a nuclear unresolved (r$<$30\,pc) HD and FC emission. The mean age map shows that this source is dominated by a population younger than 1.5\,Gyr.
  
    \item[{\bf NGC~4151:}] This galaxy displays a strong contribution of young SPCs. The $xy$ population is clearly detected on the west side of the FoV, summing up 50 percent contribution, as well as on the east side, starting at 100\,pc from the nucleus up to the edge of the FoV. A ring-like structure ($\sim$30\,pc$< r <$ 150\,pc) composed predominantly of the $xiy$ SPC, which is responsible in some locations for up to 75 percent of the emitted light, can be seen in the maps. Small fractions of $xyo$ ($\leq$ 30 percent) stars can also be found in this ring-like structure.  This component surrounds a nuclear ($r<$40\,pc) contribution from an old population. A very strong contribution of hot dust emission is observed in the unresolved nucleus ($r\lesssim$10\,pc) of this source and extends towards the full FoV. The FC component is also seen up to $r\sim$60\,pc.  With the nucleus being dominated by hot dust emission, the mean age map shows a ``hole", with a peak of $\sim$5\,Gyr SPC at 20\,pc from the centre, beyond which it has a strong negative gradient out to 40\,pc and then is nearly constant (t$\sim$500\,Myr) in the rest of the FoV. 
 
  \item[{\bf NGC~4235:}] This galaxy shows a very complex SFH over the full FoV. A (partial) ring of $xy$ is clearly present, with values reaching 50 percent to northern ($r>$\,130\,pc) and western ($r\gtrsim$100\,pc) locations, where a vertical stripe can also be seen, and which may be attributed to a residual fingerprint. Small fractions ($\leq$ 20 percent) of the $xyo$ SPC are detected in NW and SE directions ($r\gtrsim$130\,pc), while a partial ring of intermediate population ($xiy$ and $xio$) is detected towards E-S-W directions surrounding the centre (50\,pc\,$<r<$\,150\,pc). The $xo$ component shows up on the full FoV with values reaching 50 percent. A clear dominance (75 percent) of hot dust emission is seen in the unresolved nucleus ($r<$\,25pc) with a contribution of FC emission of less than 20 percent. The $HD$ component shows a steep gradient decreasing out to the edge of the FoV ($r\sim$\,200\,pc), while the $FC$ seems to reach its maximum (20 percent) for $r\sim$40\,pc towards SE directions.
   The mean age map shows the light is dominated by a 500 Myr population with a flat gradient, raising up again outwards of $\sim$200\,pc. 
  
   \item[{\bf NGC~5506:}] The two youngest SPC show a negligible contribution to the integrated spectrum emission of this galaxy. A ring-like structure (90\,pc\,$< r <$\,120\,pc) is observed for the $xiy$ component, while for the two oldest components an arc structure is observed towards the west  direction ($r>$90\,pc) with a maximum value of 25 percent for the $xo$. The dominant feature we have detected in this source is the hot dust emission, completely dominating the nuclear emission. Meanwhile, no contribution of the FC component was found in the nuclear region, although it reaches up to $\sim$50 percent in the outer regions of the FoV ($r>$90\,pc). We interpret this as due to hot dust emission instead of scattered light from the AGN. The mean age map also shows the ring signature pointing to very young ages in the centre. This, however, seems to be a fitting artefact, since the nuclear emission reaches $\sim$100 percent for the hot dust emission. 
  
    \item[{\bf NGC~5548:}] The FoV of this galaxy is dominated by the $xo$ population, with a ring ($r>$200\,pc)  of $xyo$ populations. $HD$ and $FC$ are detected along the full FoV. A detailed analysis is made by \citet{Schonell+17}.

  \item[{\bf NGC~5899:}] The two youngest components show a small contribution over the FoV, with an enhancement to the West/NW side ($\sim$ 75 percent for $r>$100\,pc in this side). The intermediate age components are clearly the dominant ones with $xiy$ being enhanced on the East side of the FoV (where it reaches a $\sim$75 percent contribution). The $xio$ SPC is well distributed over the FoV with a mean value of $\sim$40 percent in the inner 100\,pc radius. For the old SPC we can see a flat distribution over the entire FoV, with a contribution of $\sim$20 percent. Very small fractions of $HD$ and $FC$ are required in the central region and the mean age shows a small gradient with values dropping from $\sim$1.3\,Gyr to $\sim$ 0.7\,Gyr in the outer regions.

\item[{\bf NGC~5929:}] A very small fraction of $xy$ SPC is observed SE of the nucleus ($r>$110\,pc) of NGC~5929, while no signs of the $xyo$ component are seen. The dominant emission of the central region is due to the $xiy$ SPC reaching 75 percent in most locations of the FoV, and presenting a value of 25 percent in the unresolved nucleus ($r<$\,25\,pc), which is dominated by the $xio$ population (up to 50 percent). No hot dust and $FC$ emission are detected. The mean age shows a decreasing gradient with ages between 1.5~Gyr in the nucleus and 0.6~Gyr in the outermost locations ($r\sim$\,100\,pc). 
  
  \item[{\bf Mrk~607:}] The SFH of the central region of this galaxy is very complex and all the SPCs are necessary to explain the continuum emission of this galaxy. The two youngest SPCs ($xy$ and $xyo$) when taken together present a partial ring, from south towards west up to north (130\,pc\,$< r < $\,150\,pc). The $xiy$ SPC shows another ring-like structure (100\,pc\,$<r<$150\,pc), reaching up to 50 percent of the emission. The $xio$ SPC is spread over the FoV, and summed with $xo$ contributes up to 40 percent of the stellar emission in the nuclear region ($r<$50\,pc). The clear dominant nuclear emission in this galaxy is due to the hot dust emission, reaching more than 50 percent in the centre, and falling to zero in the outer locations. A tiny fraction ($<$ 5 percent) of a $FC$ is also detected in the unresolved nucleus. The mean age shows a decreasing gradient, with a peak of $\sim$2.5~Gyr in the centre falling to values $\lesssim$ 1~Gyr in the outer locations.
  
  \item[{\bf Mrk~766:}] This source also shows a very complex SFH, with a ring of very young ($xy$) stars dominating the emission in the outer regions of the FoV ($r\gtrsim$200\,pc). In this same ring there is a coexistence of all other SPCs, in smaller fractions. The centre is dominated by hot dust ($\sim$60 percent) and $FC$ ($\sim$ 30 percent) emission. The mean age map indicates that slightly older stars are located in a SE direction. 

   \item[{\bf Mrk~1066:}] The FoV of this galaxy is dominated by the $xiy$ population. $HD$ and $FC$ emission is clearly observed in the unresolved nucleus, decreasing outwards over the full FoV. A detailed analysis is made by \citet{Riffel+10}.
   
   \item[{\bf Mrk~1157:}] The FoV of this galaxy is dominated by the intermediate age populations ($xiy$ and $xio$) with a smaller contribution  of $xo$. Small fractions of $HD$ and $FC$ emission are observed in the unresolved nucleus. A detailed study of the stellar populations for this galaxy is made in \citet{Riffel+11}.
   
\end{description}

\begin{figure*}
    \centering
    \includegraphics[scale=0.7]{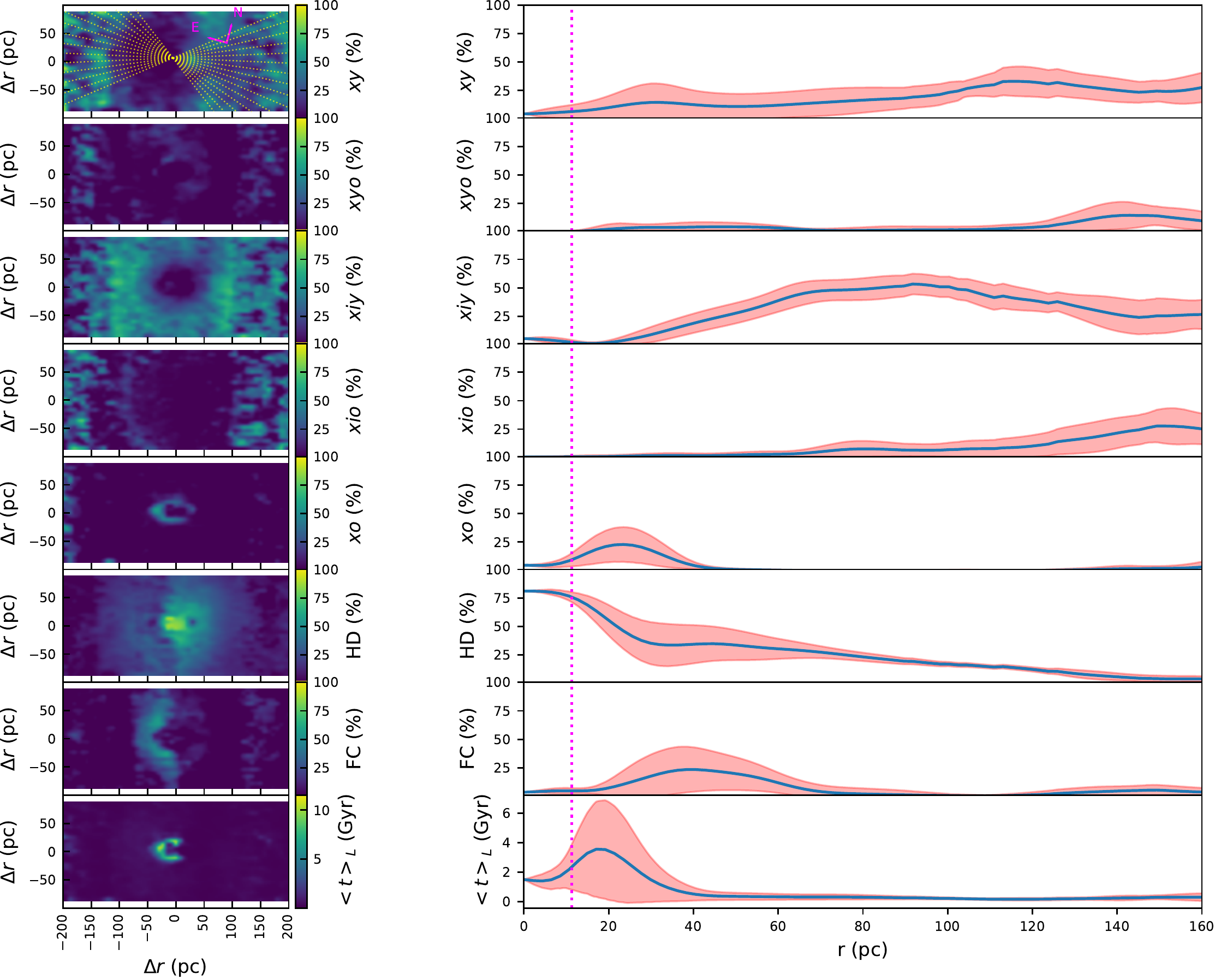}
    \caption{Maps and radial profiles for NGC\,4151. Left side, from top to bottom: 
      $xy$ percent  contribution of SSPs with ages $t\leq$ 50~Myr;
     $xyo$ percent contribution of SSPs with ages  in the range 100~Myr $\leq t \leq$ 200~Myr;
     $xiy$  contribution of SSPs with ages  in the range 500~Myr $\leq t \leq$ 700~Myr;
    $xio$  contribution SSPs with ages  in the range 1~Gyr $\leq t \leq$ 2~Gyr;
    $xo$  contribution 5~Gyr $\leq t \leq$ 13~Gyr;
    $HD$   contribution of all Plank functions components;
    $FC$ is the featureless continuum percent contribution;
    $<t>_L$ is the light weighted mean age. In the right side we show the radial profiles (yellow dotted lines on the first panel) showing the azymuthal mean of the corresponding properties, together with their standard deviation (shaded region). The dotted vertical line is the PSF of the FWHM taken from \citet{Riffel+17,Riffel+21a}. We have identified spaxels with a percentage mean deviation ($AdevS$) larger than 10 (red contour) and 20 (white contour) percent. See text for more details.} 
    \label{maps0}
\end{figure*}

\begin{figure*} 
 \centering 
  \includegraphics[scale=0.6]{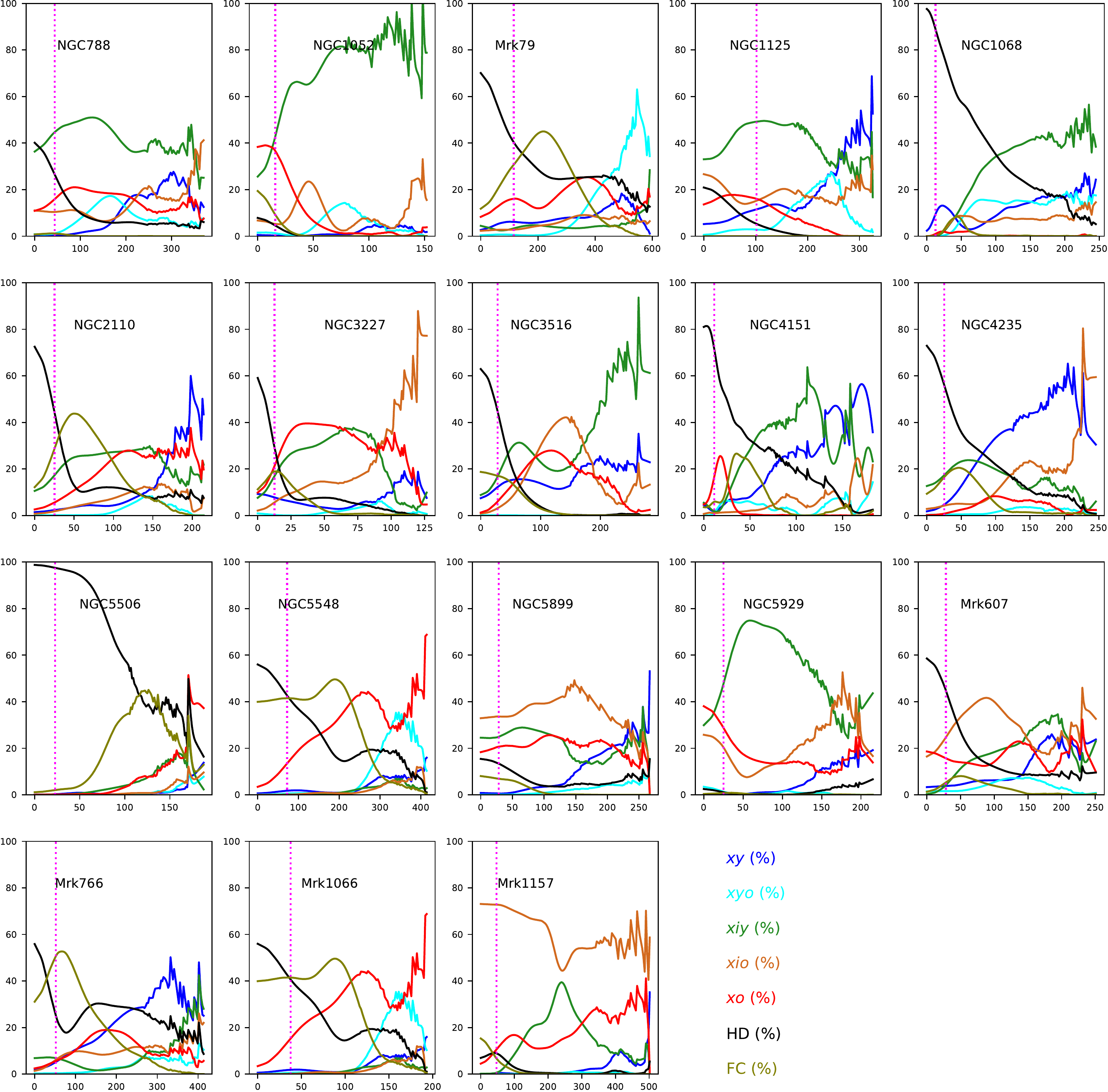} 
 \caption{ Individual profiles for each population vector. 
 The dotted vertical line is the PSF of the FWHM taken from \citet{Riffel+17,Riffel+21a}. It can be seen that the SFH of these sources if complex. The $HD$ component dominates the emission of the unresolved nucleus for almost all the sources, while the $FC$ reaches its maximum value for some galaxies outside the nucleus.
 Very high fractions of young to intermediate age stars are observed in regions with ($r \gtrsim$ 100\,pc). 
 See text for more details.} 
 \label{indpropProfs}
\end{figure*}

To summarise the individual results, we have plotted in Fig.~\ref{indpropProfs} the individual profiles of the binned population vectors. These figures show the results for all the 18 galaxies, including the 6 objects previously studied by our group. It is clear from this figure that the inner region has centrally peaked hot dust emission in the unresolved region (dotted line). On the other hand, for the $FC$ contribution we found that for nearly half of our sample this component reaches the highest values outside the unresolved nucleus (see \S~\ref{hotdust}).

In terms of stellar populations, the galaxies show a considerable fraction of young ($xy$+$xyo$) stellar populations; however, the central region is dominated by an intermediate age population ($xiy$+$xio$), as well as a significant fraction of old stars. In general, the galaxies display a very complex star formation history in the central region.

\subsection{Sample Results}

In order to compare the stellar population results for our sample with the properties listed in Tab.~\ref{tab_obs} we have followed \citet[][]{Riffel+21a} and have computed the mean values of the derived stellar population properties inside a 125\,pc radius. These values are listed in Tab.~\ref{tab_derivedProps}. It is worth mentioning that since we are taking the mean value over the 125\,pc in centre of the galaxies, the possible fingerprints left on the borders of the FoV do not play an important role on the obtained values.

\begin{table*}
\renewcommand{\tabcolsep}{0.6mm}
\centering
\small
\caption{Mean values of the derived stellar population properties within the inner 125\,pc.}
\vspace{0.3cm}
\begin{tabular}{lllllllllllll}
\hline
\hline
Source         &          $xy$        &        $xyo$         &        $xiy$         &       $xio$          &        $xo$          &       $HD$           &        $FC$           &      $<t>_{L}$    &        $<t>_{M}$    &     $<Z>_{L}$       &     $<Z>_{L}$       &     $Av$             \\
\noalign{\smallskip}
\hline	
NGC\,788       &	   2.23$\pm$4.72  & 	  8.9$\pm$13.54  &     48.77$\pm$19.72  &	  15.52$\pm$19.77  &	 15.66$\pm$11.6   & 	  8.9$\pm$10.53  &  	 0.0$\pm$0.02	 &    0.94$\pm$0.43  &  	3.21$\pm$2.14  &	  0.03$\pm$0.01  &  	0.02$\pm$0.01  &	  0.41$\pm$0.39  \\  
NGC\,1052      &	   3.28$\pm$7.77  & 	11.36$\pm$14.84  &     73.18$\pm$21.53  &	   7.66$\pm$17.77  &	  3.87$\pm$ 9.8   & 	 0.21$\pm$1.35   &  	0.44$\pm$3.43	 &     0.6$\pm$0.46  &  	1.27$\pm$1.75  &	  0.03$\pm$0.01  &  	0.03$\pm$0.01  &	  0.11$\pm$0.45   \\ 
Mrk\,79        &	   5.43$\pm$7.45  & 	19.08$\pm$29.08  &  	8.38$\pm$10.92  &	   2.91$\pm$3.49   &	 11.55$\pm$14.26  & 	28.45$\pm$23.73  &     24.21$\pm$27.1	 &    1.01$\pm$1.14  &  	3.61$\pm$3.21  &	  0.02$\pm$0.01  &  	0.01$\pm$0.01  &	   2.0$\pm$1.36   \\ 
NGC\,1125      &	   15.9$\pm$17.19 & 	 7.57$\pm$11.05  &     48.36$\pm$29.79  &	  15.89$\pm$18.44  &	  7.87$\pm$13.34  & 	 4.42$\pm$6.03   &  	 0.0$\pm$ 0.0	 &    0.57$\pm$0.55  &  	 1.9$\pm$2.83  &	  0.03$\pm$0.01  &  	0.03$\pm$0.01  &	  0.08$\pm$0.23   \\ 
NGC\,1068      &	  10.66$\pm$7.04  & 	13.05$\pm$8.37   &     29.47$\pm$16.41  &	  10.67$\pm$12.11  &	  0.77$\pm$1.94   & 	34.17$\pm$19.21  &  	1.22$\pm$3.14	 &    0.29$\pm$0.16  &  	0.64$\pm$1.23  &	  0.02$\pm$0.01  &  	0.02$\pm$0.01  &	  0.25$\pm$0.74   \\ 
NGC\,2110      &	  12.35$\pm$14.77 & 	 1.64$\pm$5.08   &     24.04$\pm$10.58  &	   7.94$\pm$9.47   &	  19.2$\pm$13.33  & 	16.11$\pm$15.28  &     18.72$\pm$17.27   &    0.97$\pm$0.75  &  	5.41$\pm$2.51  &	  0.02$\pm$0.01  &  	0.02$\pm$0.01  &	   1.5$\pm$0.79   \\ 
NGC\,3227      &	    5.4$\pm$9.18  & 	 2.49$\pm$6.48   &     28.21$\pm$21.07  &	  27.29$\pm$27.52  &	 28.87$\pm$16.73  & 	 5.45$\pm$8.54   &  	2.28$\pm$6.19	 &    1.56$\pm$0.93  &  	5.13$\pm$2.51  &	  0.02$\pm$0.01  &  	0.02$\pm$0.01  &	  1.95$\pm$0.69   \\ 
NGC\,3516      &	  15.85$\pm$14.19 & 	 0.02$\pm$0.13   &     30.79$\pm$25.01  &	  21.64$\pm$18.96  &	 22.59$\pm$15.73  & 	 5.68$\pm$15.79  &  	3.44$\pm$11.53   &    0.92$\pm$0.71  &  	4.97$\pm$2.81  &	  0.01$\pm$0.01  &  	0.01$\pm$0.01  &	  1.29$\pm$0.94   \\ 
NGC\,4151      &	  37.39$\pm$28.46 & 	 2.76$\pm$5.66   &     25.36$\pm$22.26  &	   9.12$\pm$13.8   &	  5.12$\pm$12.03  & 	16.54$\pm$17.17  &  	3.71$\pm$8.93	 &    0.39$\pm$1.25  &  	1.62$\pm$3.08  &	  0.02$\pm$0.01  &  	0.02$\pm$0.01  &	  0.16$\pm$0.73   \\ 
NGC\,4235      &	  29.84$\pm$23.03 & 	 2.38$\pm$5.97   &     21.93$\pm$17.79  &	  10.95$\pm$13.6   &	  4.77$\pm$7.93   & 	22.14$\pm$19.48  &  	7.98$\pm$13.42   &    0.33$\pm$ 0.4  &  	1.93$\pm$2.19  &	  0.02$\pm$0.01  &  	0.02$\pm$0.01  &	   2.1$\pm$1.06   \\ 
NGC\,5506      &	   1.06$\pm$2.67  & 	  0.7$\pm$4.16   &  	8.94$\pm$10.36  &	   2.93$\pm$6.21   &	  6.78$\pm$11.57  & 	56.83$\pm$36.61  &     22.76$\pm$26.14   &    1.37$\pm$1.83  &  	3.23$\pm$3.46  &	  0.01$\pm$0.01  &  	0.01$\pm$0.01  &	  1.83$\pm$2.05   \\ 
NGC\,5548      &	    5.9$\pm$8.85  & 	 9.54$\pm$16.12  &  	2.39$\pm$ 8.5	&	   2.82$\pm$8.68   &	  26.7$\pm$21.34  & 	25.13$\pm$24.12  &     27.51$\pm$28.59   &    3.95$\pm$3.75  &  	6.62$\pm$3.12  &	  0.01$\pm$0.01  &  	 0.0$\pm$ 0.0  &	  2.96$\pm$1.78   \\ 
NGC\,5899      &	  14.84$\pm$23.44 & 	 4.89$\pm$10.73  &     25.27$\pm$26.85  &	  28.96$\pm$25.96  &	  19.1$\pm$14.53  & 	 5.41$\pm$7.83   &  	1.53$\pm$4.12	 &    1.11$\pm$0.86  &  	 4.5$\pm$ 2.8  &	  0.02$\pm$0.01  &  	0.01$\pm$0.01  &	   1.7$\pm$0.85   \\ 
NGC\,5929      &	   2.71$\pm$6.09  & 	 3.25$\pm$10.14  &     61.07$\pm$25.42  &	   17.4$\pm$23.7   &	 14.42$\pm$13.33  & 	 0.82$\pm$1.81   &  	0.31$\pm$1.64	 &    0.94$\pm$0.48  &  	3.32$\pm$ 2.2  &	  0.02$\pm$0.01  &  	0.01$\pm$0.01  &	  0.61$\pm$0.71   \\ 
Mrk\,607       &	  13.51$\pm$17.54 & 	 5.56$\pm$10.75  &     20.99$\pm$18.78  &	  27.31$\pm$21.68  &	 14.39$\pm$15.39  & 	 15.3$\pm$11.83  &  	2.94$\pm$5.61	 &    1.23$\pm$1.07  &  	3.78$\pm$3.15  &	  0.03$\pm$0.01  &  	0.02$\pm$0.01  &	  1.45$\pm$0.83   \\ 
Mrk\,766       &	  11.51$\pm$10.5  & 	 2.94$\pm$7.15   &  	4.44$\pm$8.54	&	  18.05$\pm$18.5   &	  15.3$\pm$15.16  & 	25.87$\pm$20.03  &     21.88$\pm$21.12   &    1.23$\pm$1.35  &  	5.91$\pm$3.39  &	  0.03$\pm$0.01  &  	0.02$\pm$0.01  &	  2.02$\pm$1.08   \\ 
Mrk\,1066      &	  18.88$\pm$30.18 &       1.1$\pm$2.23   &      61.55$\pm$31.87 &      6.75$\pm$13.09  &      4.85$\pm$7.17   &      5.96$\pm$ 9.7   &      0.91$\pm$2.11    &    0.47$\pm$0.3   &      1.45$\pm$1.29  &      0.03$\pm$0.01  &      0.03$\pm$0.01  &      0.35$\pm$0.48   \\ 
Mrk\,1157      &	   5.65$\pm$10.28 & 	 0.42$\pm$1.97   &     19.71$\pm$26.67  &	   55.3$\pm$31.6   &	 15.48$\pm$14.83  & 	 2.03$\pm$3.02   &  	1.41$\pm$ 3.4	 &    1.63$\pm$0.92  &  	3.52$\pm$1.92  &	  0.03$\pm$ 0.0  &  	0.03$\pm$0.01  &	  0.22$\pm$ 0.3   \\ 
\noalign{\smallskip}
\hline
\end{tabular}
\label{tab_derivedProps}
%\begin{list}{Table Notes:}
%\item (1) Galaxy's name; 
%\end{list}
\end{table*}

\begin{figure*} 
 \centering 
  \includegraphics[width=\textwidth]{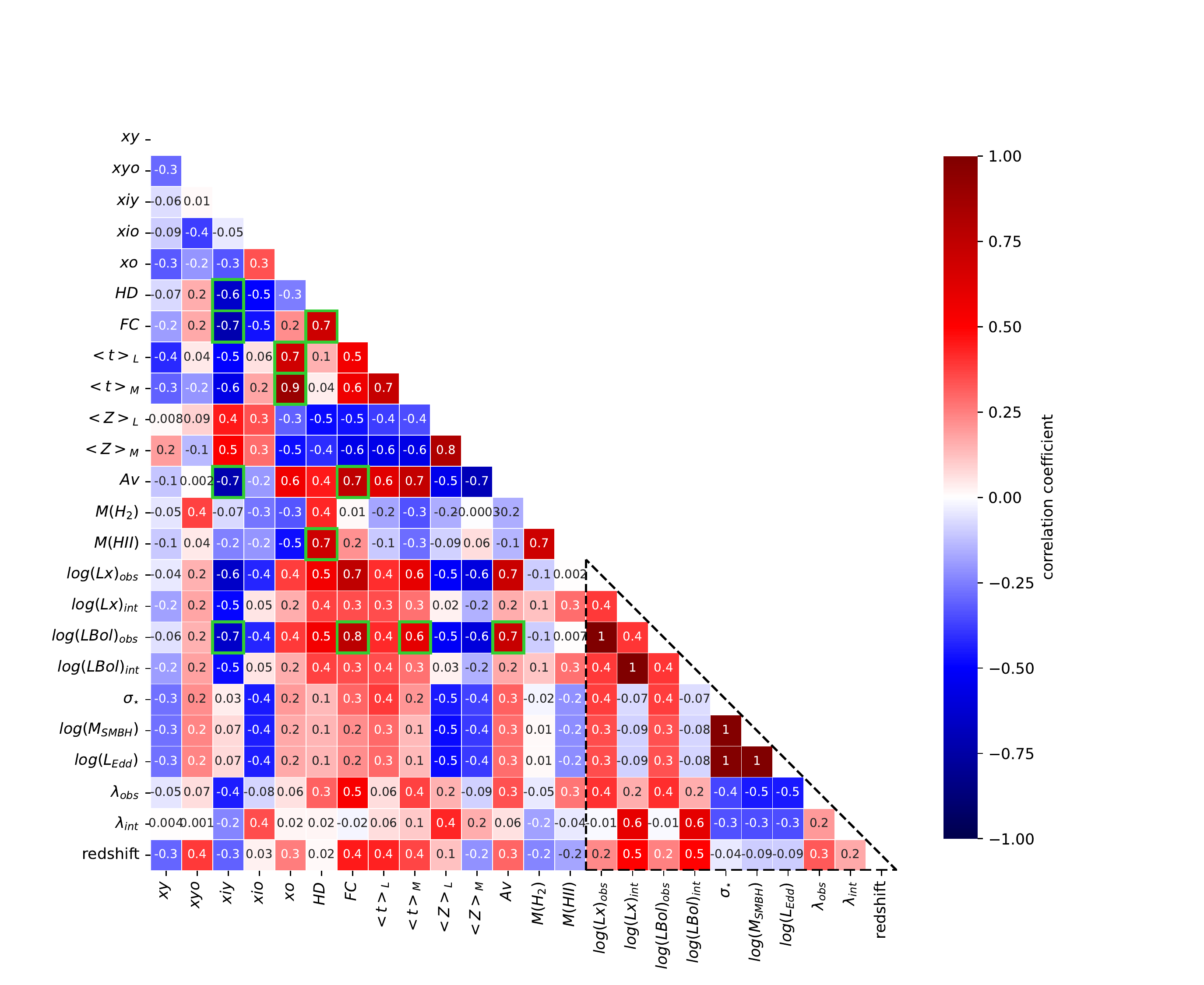} 
 \caption{Diagonal Pearson standard correlation coefficient matrix comparing the sample properties and derived stellar population properties. The triangle represented with dashed lines show the coefficients of properties obtained from literature or which derivation are dependent with each other. The meaningful correlations are plotted in Fig.~\ref{indCor} and are highlighted with green boxes. For details see text.}
 \label{CorMatrix}
\end{figure*}

When merging Tab.~\ref{tab_obs} and  Tab.~\ref{tab_derivedProps} data together we are left with a large set of properties. To compare all the properties we have computed the Pearson standard correlation coefficient matrix using the {\sc seaborn}  statistical data visualisation Python package \citep{Waskom+21}. The diagonal correlation matrix is shown in Fig.~\ref{CorMatrix} where the triangle traced by dashed lines shows the correlation coefficients of properties obtained by derivations that may depend on each other ($log(Lx)_{obs}$,  $log(Lx)_{int}$, $log(LBol)_{obs}$, $log(LBol)_{int}$, $\sigma_{\star}$, $log(M_{SMBH})$, $log(L_{Edd})$, $\lambda_{obs}$, $\lambda_{int}$ and redshift) and therefore will not be discussed here, but are left in the figure for completeness. 

\subsection{Caveats on hot dust {\it versus} featureless continuum: possible degeneracies}\label{hotdust}

Our analysis points towards the detection of $HD$ and/or $FC$ outside the unresolved nucleus in nearly half of the sources (NGC\,4151, Mrk\,79, NGC\,3227, NGC\,3516, NGC\,4235, NGC\,5506, Mrk\,766, NGC\,1068, NGC\,2110, NGC\,5548 and Mrk\,1066). 

The case of the $FC$ component is a known and common problem in the study of the stellar content of Seyfert galaxies. It is related to the fact that the continuum of a reddened young starburst (t $\lesssim$ 5\,Myr) is indistinguishable from
an AGN-type continuum, which is why we have not added such young (t\,$<$\,10\,Myr) components in the base of elements \citep[for more details see][and references therein]{Riffel+09}. This could be used to explain the fact of the $FC$ component showing up outside the unresolved region, in the case of regions dominated by the very young populations (J or J+H bands dominated by young stellar populations). In this case, this fraction would no be a true AGN component, but due to a very young reddened stellar population.

As can be seen from Fig.~\ref{base} a degeneracy between $FC$ and $BB$ components can occur when the hot dust (or FC) dominates the K-band emission and the stellar population dominates the J-Band (or J+H). In that case, the main  distinguishing characteristic will be the slope of the continuum in the K-band. In this case, the $FC$ component is hardly distinguishable from dust with $T\gtrsim 1200$K.  In contrast, in situations where either stellar light dominates all three  bands (i.e. away from the nucleus), or the $HD$+$FC$ dominate all bands (right on the nucleus), there will be no issue. The  degeneracy will only show up in the transition regions. In order to test this, we performed a fit in which we only included the $FC$ component in the central ($r$=2 pixels) unresolved region. For the
cases where there had been some fraction of $FC$ in the outer region, this was now redistributed among the $BB$ components, with very little going into the SSP components.

Complementing these tests is the fact that the $log(LBol_{obs})$ and $FC$ correlation is only strong for larger fractions of $FC$. We interpret this as implying that when there is a significant $FC$ fraction ($\ge$ 15 percent), it can be associated with the accretion disc emission. This result is in agreement with the fact that broad components in polarised light are associated with larger fractions of the $FC$ component \citep{CidFernandes+95,CidFernandes+04,Riffel+09}. 

These results, taken together with the fact that hot dust emission outside the unresolved region has been previously reported in AGNs \citep[e.g.][]{Davies+05,Dottori+05,Gratadour+06,Martins+10a,Storchi-Bergmann+12a, Gaspar+19}, makes us to interpret the $FC$ detected in the transition regions (outside the unresolved region) as due to hot dust emission  (which, in that case, would be heated by young stellar populations) or to a reddened very young stellar population, and not to a true AGN $FC$ component.

 \section{Discussion} \label{discussion} 

\begin{figure*} 
 \centering 
  \includegraphics[scale=0.65]{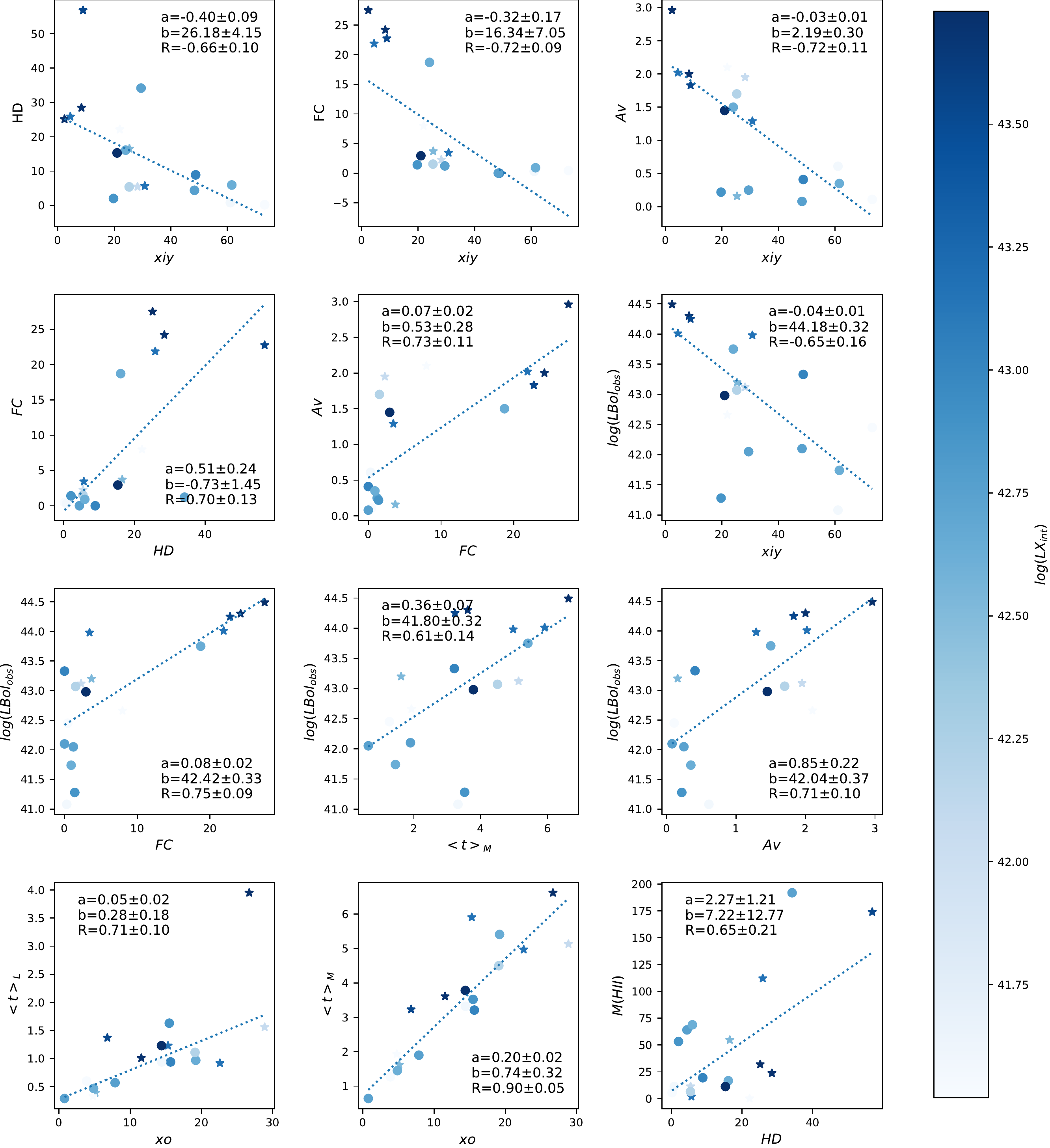} 
 \caption{Correlations of different properties. (stars represent Sy~1 and circles Sy~2). The uncertainties on the a linear fit (dotted blue line, y=ax+b) to the data points (circles) was obtained using 1000 bootstrap realisations with Huber Regressor model. The Pearson's correlation coefficient ($R$), slope (a) and intercept (b) are labelled. The error bars for a and b, are the standard deviation of the mean after the bootstrap realisations. We have only considered as statistically significant correlations were the difference between the absolute mean r value minus the standard deviation was larger than 0.45 ($|R| - \delta R >$0.40). The blue dotted line is obtained with the mean a and b values. The logarithmic of the X-ray intrinsic luminosity is represented by different colours. For details see text.}  
 \label{indCor}
\end{figure*}

When inspecting the correlation matrix (Fig~\ref{CorMatrix}) we see that several quantities are well correlated (|R|$>$0.5). After removing the possible correlations with $LX_{obs}$, since they should be the same as those with $LBol_{obs}$ (that was derived using $LX_{obs}$), we have made scatter plots for all potential correlations (|r|$>$0.5). These are shown in Fig.\,\ref{indCor} together with a linear fit to the data points (dotted blue line) obtained using bootstrap realisations \citep{Davison+97} with the Huber Regressor model that is robust to outliers \citep{Owen+07}. The Pearson's correlation coefficient ($R$), slope (a) and intercept (b) are also quoted. The uncertainties on the correlation coefficients are the standard deviation of the mean after 1000 bootstrap realisations. We have only considered as statistically significant correlations where the difference between the modulus of the mean $R$ value minus the standard deviation was larger than 0.45 ($|R| - \delta R >$0.45). From this we got the following meaningful correlations:
 $xiy \times HD$; 
 $xiy \times FC$;
 $xiy \times Av$;
 $xiy \times log(LBol_{obs})$;
$xo \times <t>_L$;
 $xo \times <t>_M$;
 $HD \times M(HII)$;
 $HD \times FC$;
 $FC \times Av$;
 $FC \times log(LBol_{obs})$;
 $<t>_M \times log(LBol_{obs})$ and 
 $Av \times log(LBol_{obs})$.

 A clear anti-correlation of the intermediate young ($xiy$; 500~Myr $\leq t \leq$ 700~Myr) stellar population component with the two featureless AGN components (HD and FC) and reddening is observed. This is further confirmed by the correlation of $FC$ and $HD$ as well as $Av$ and $FC$, since these quantities are individually related with $xiy$.

In addition, $xiy$ is also anti-correlated with the logarithm of the observed bolometric luminosity and there is a trend in the sense that higher intrinsic X-ray luminosity is observed for higher fractions of $HD$, $FC$ and $Av$ (see Fig~\ref{indCor}).

 The correlation between the X-ray luminosity (seen via the correlations with $LBol$) and $HD$, $FC$ and $A_V$ can be tentatively attributed to the fact that all these properties are linked to the mass accretion rate onto the AGN, as follows.  More $A_V$ means more gas, thus more fuel to the AGN, and more fuel results in increased accretion rate. This in turn leads to higher luminosity of the accretion disk, thus increased contribution of the $FC$ continuum. This continuum is the source of the heating of the circumnuclear dust, and thus the $HD$ contribution also increases. A closer look at these plots (third row of panels in Fig.\,\ref{indCor}) shows that the correlation is dominated by the most luminous sources ($log(LBol)\ge$43.5) and can be understood as due to the fact that, for the low luminosity end, secondary parameters may play an important role \citep[see][for example]{Yang+15}.

The anti-correlation of the above properties with $xiy$ may be a secondary effect, resulting from the fact that all of them are correlated, meaning that, when all their contributions increase, the result is a decreased contribution of the stellar population.

The mean ages (light and mass weighted) are well correlated with the $xo$ component and a tighter correlation is observed between $xo$ and $<t>_M$, due to the non-linear behaviour of the M/L fractions of stellar populations \citep[see][for additional discussion]{Riffel+09}. However, it is important to note that the fractions of $xo$ are less than 30 percent and that the mean age is actually dominated by the younger components (see Eq.~\ref{ewMage}). 

$HD$ is correlated with the mass of \ion{H}{ii} and with the $FC$ fraction, suggesting that the hot dust present in the inner region of these galaxies is heated by the AGN $FC$. This is in agreement with the previous interpretation: more gas means increased accretion rate, increasing the $FC$, which increases the heating of the dust, resulting in a larger $HD$ fraction.

As expected $log(LBol_{obs})$ is well correlated with the $FC$ fraction, since the bolometric luminosity is primarily driven by the AGN accretion disk emission. This correlation is very strong for $FC \ge$ 15 percent, while for small values of $FC$ a scatter is observed (see \S~\ref{hotdust}).

A correlation of $log(LBol_{obs})$ and $<t>_M$ is also observed. This correlation remains when, instead of the mass-weighted mean ages, we use the light-weighted ones. However, it is weaker, but in the case of light weighted mean ages the values for almost all sources are $<t>_L \lesssim $ 1.5\,Gy. We interpret this correlation as due to a delay between the formation of new stars and the triggering/feeding of the AGN. The cold gas that reaches the SMBH most probably is originated from the mass-loss from intermediate age stars, as proposed by  \citet{Davies+07a} who found that star formation inhibits accretion, and that gas accretion into the SMBH is only efficient after these early turbulent phases of stellar feedback. The gas released by this intermediate age population has a low velocity (a few hundred \kms) and is accreted together with the gas already available in the central region \citep[e.g.][]{Cuadra+06,Hopkins+12}. This extra amount of gas will trigger the AGN (or make it brighter).
The intermediate age population is dominated by C- and O-rich stars  \citep[e.g.][]{Maraston+05,Dottori+05,Riffel+07,Riffel+15,Salaris+14a}, and these short lived stars ($t\simeq $0.2 -- 2\,Gyr; $\rm M\simeq 2 - 6\,M_{\odot}$) do release enriched material to the nuclear environment (fuelling the AGN). This material would remain, held back by the central potential, and may explain the $HD$ that is observed outside the unresolved nucleus.

 We also compared our findings with the optical stellar population results of \citet{Burtscher+21}. They analyzed the inner 150\,pc of nine AGN hosts from a complete volume-limited sample of targets with $L_{14-195keV}>10^{42.5}$ from the Local Luminous AGN with Matched Analogs (LLAMA) sample. Out of their nine AGNs, they detected young SPs in seven. On average, the light contribution\footnote{It is worth mentioning that the comparison of optical and NIR young stellar populations is not direct \citep{Riffel+11c}.} from these young stellar populations was 4.5\%, compared to 11.8\% in our sample.  Another difference between our results and theirs is the fact that in all their AGNs except one, the light is dominated by old SPs, whereas this is the case for only two of our galaxies (NGC\,3227 and NGC\,5548). A similar analysis was done by \citet{Dahmer-Hahn+21}, who analyzed the stellar populations using optical IFU data of 14 AGN (being the optical counterpart of the present sample). Similarly to the results of \citet{Burtscher+21}, nine of their objects have evidence of young SPs in the inner 150\,pc, with an average light contribution of 7~\%. Also, their sample is also dominated by old SPs, with more than 70~\% of the light coming from SPs older than 2\,Gyr stellar populations. 
\par
Compared to these optical works, we detected young and intermediate-age SPs in a higher fraction of galaxies, and with a higher average contribution. This is a direct consequence of the lower reddening effects in the NIR, meaning that we can access SPs which are inaccessible in the optical due to dust obscuration, together with the fact that the NIR is more sensitive to the colder bright stellar phases \citep{Maraston+05,Salaris+14a,Riffel+11a,Riffel+11c,Riffel+15}.

\begin{figure*} 
 \centering 
  \includegraphics[width=\textwidth]{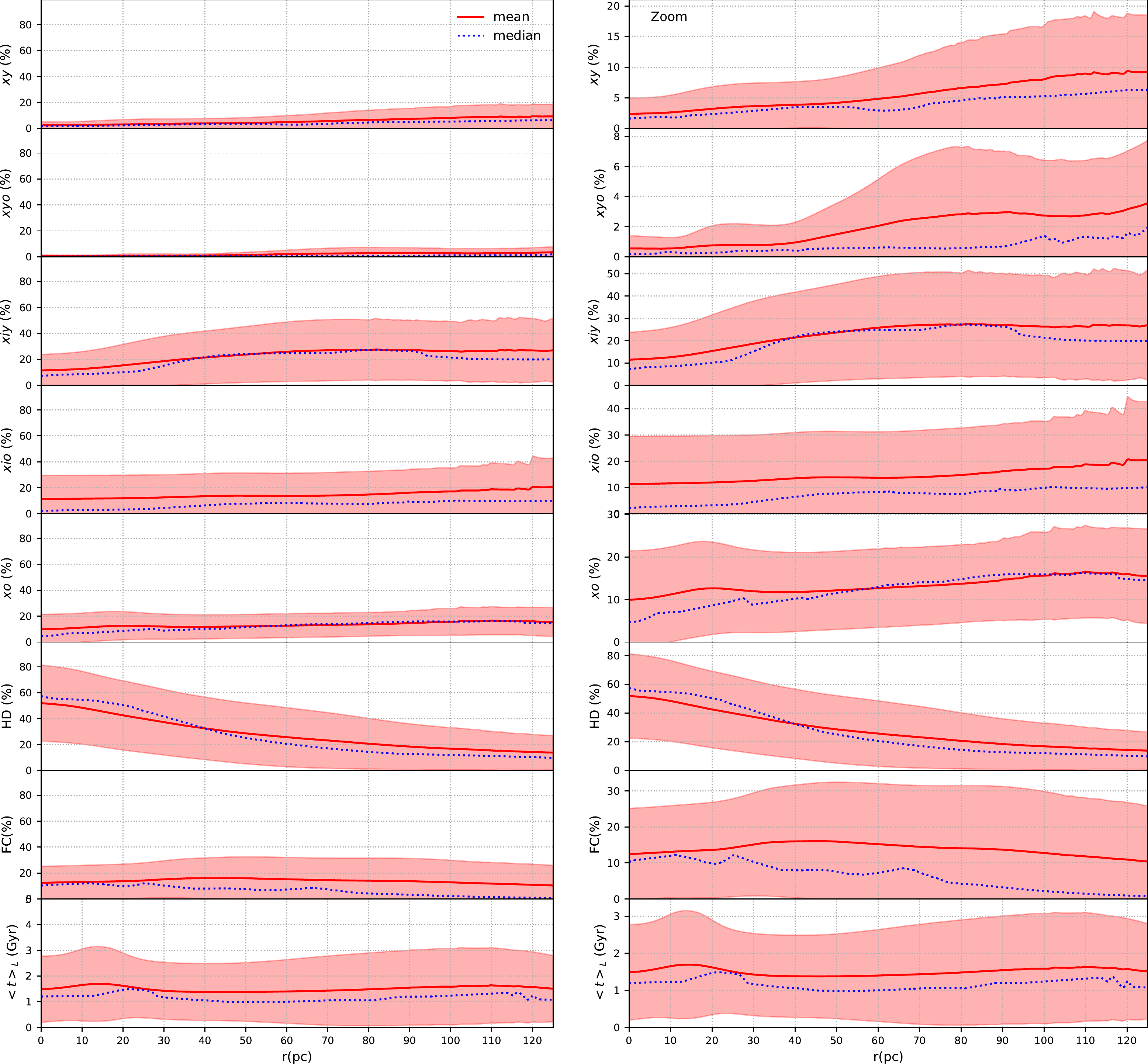} 
 \caption{Mean and median AGN profiles of different properties (left to the same scale and right a zoom in). These profiles where obtained as the mean/median values of the individual profiles for each AGN (see Fig.~\ref{maps0} and the supplementary material, for example). The red shaded regions are the standard deviation. It is clear from that figure that the inner pc of the galaxies are dominated by an intermediate age stellar population. } 
 \label{MeanProfiles}
\end{figure*}

As discussed above, in Fig.~\ref{indCor} we have obtained a mean value of the stellar population properties for the inner 125\,pc of each galaxy. In order to better understand the variation of these  properties as a function of distance to the nucleus, we have computed mean and median profiles inside this region (which is the full FoV in the case of NGC3227) for the full sample. The resulting profiles are presented in Fig.~\ref{MeanProfiles}, as well as the corresponding standard deviations.

What emerges from this exercise is that the inner few parsecs (r $\lesssim$40 pc) of the galaxies have their emission dominated by hot dust emission ($HD$) and in terms of stellar content they are dominated  by intermediate age stellar populations ($xiy$ and $xyo$). This is reflected by the $<t>_L$ profile, which is nearly constant ($\sim$ 1\,Gyr) over the inner 125\,pc, reaching a maximum value of $\sim$2\,Gyr.

\begin{figure} 
 \centering 
  \includegraphics[scale=0.6]{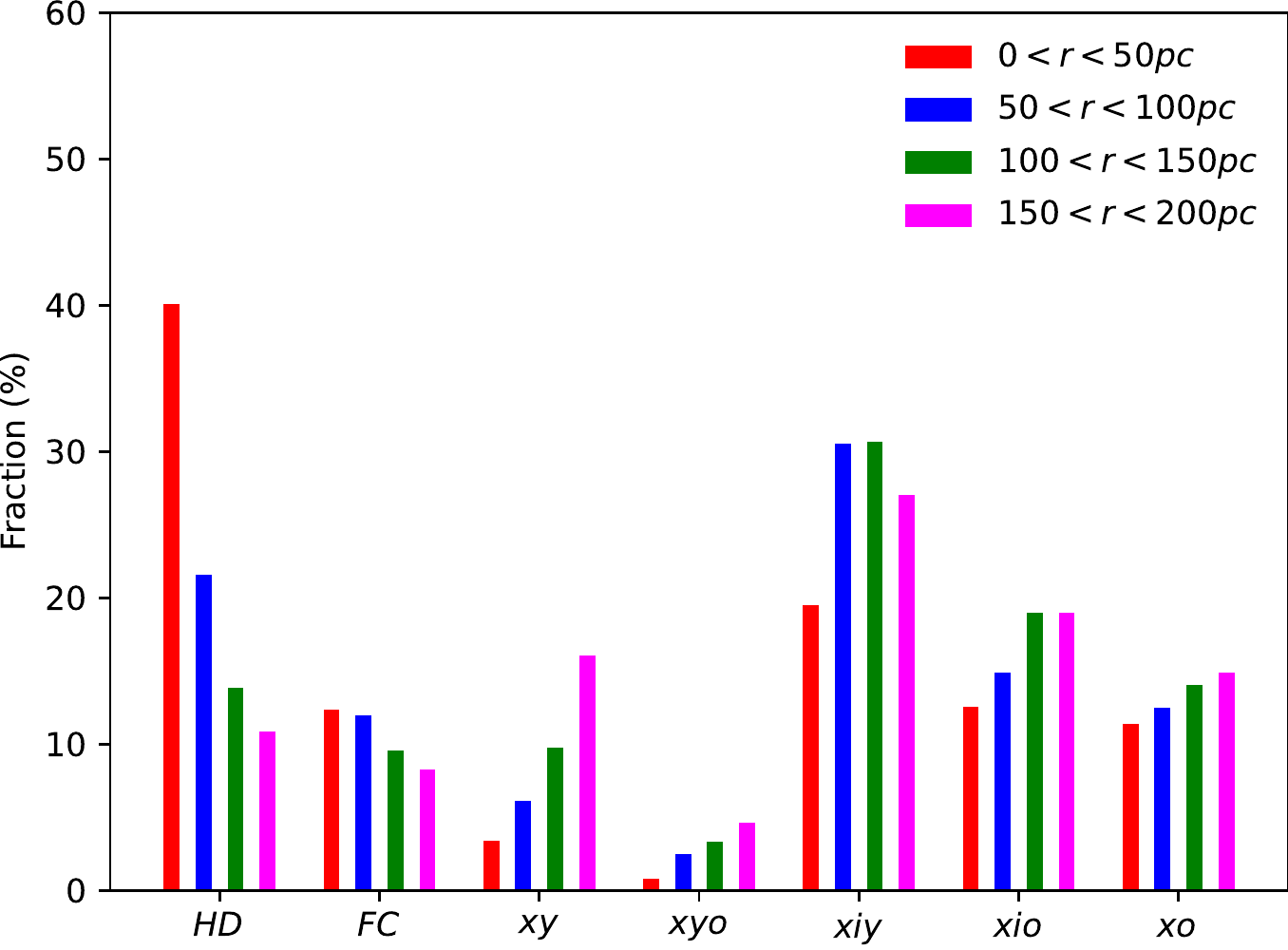} 
 \caption{Histogram of average population vectors (labelled) for different radial bins (see labels). $HD$ and $FC$ components decrease outwards, while $xy$ and $xyo$ increase outwards. The intermediate population dominates the emission over the full FoV with a trend of increasing outwards. $xo$ is nearly constant over the full FoV.} 
 \label{averagepops}
\end{figure}

In order to better understand the radial behaviour of the SFH over our sample we have produced mean values for the SPCs of our sample in four radial bins\footnote{The mean value over the profiles shown in Fig.~\ref{MeanProfiles} inside the four regions.}. The result of this procedure is shown in Fig.~\ref{averagepops}, where we show the mean values for the $HD$, $FC$, $xy$, $xyo$, $xiy$, $xio$, and  $xo$ for the different radial bins (in red for $r<$ 50 pc, in blue for 50\,pc$< r < $ 100\,pc, in gree for 100 \,pc$< r < $ 150\,pc, and in magenta for 150\,pc$< r < $ 200\,pc.). What emerges from this plot is that the $HD$  and $FC$ components decrease outwards; $xy$ and $xyo$ increase from small to larger scales; while $xiy$, $xio$, and $xo$ are nearly constant over the FoV (showing a small trend to increase outwards). 

We interpret this result as the fact that the AGN is impacting on the recent star formation, in the sense that it can be associated with the decrease of young star formation towards the nuclear region. Another interesting point is that all the components older than $xio$ (t$\gtrsim$500\,Myr) are nearly constant over the FoV. We speculate that this maybe because any episodic nature of the star formation rate is removed on those longer timescales. In particular, especially for $xio$ and $xo$, we are looking at bulge stars along the line of sight which one would not expect to be affected by what is going on in the disk plane at small scales.

When inspecting the $FC \times HD$ panel of Fig.~\ref{indCor} we observe that on one hand the correlation becomes weaker for higher factions of $FC$ (the same regime where $log(LBol_{obs})$ and $FC$ shows a tight correlation). On the other hand, for the case of low $FC$ ($\lesssim 5$ percent), $HD$ tends to be low too. As can be seen in Fig.~\ref{indCor} the majority of these points cover a wide range of $Lx$ and $Av$. We suggest that in these cases either the luminosity is low or they are very obscured, so that no $FC$ leaks out and even the $HD$, in these central regions ($r< 125$\,pc), is hidden. 

This is further supported by the fact that there is only a small correlation between $HD$ and $log(LBol_{int})$  (Fig.~\ref{CorMatrix}). We interpret this as due to the inclusion of both type one and type two sources in our analysis, and the fact that $HD$ emission is more common in Sy~1 than in Sy~2s \citep{Riffel+09}. In addition \citet{Burtscher+15}  showed that there is a good correlation between $HD$ and AGN luminosity (Fig 9 of that paper) for Sy~1, but the Sy~2 are offset by about an order of magnitude with a larger scatter. They associate this 
with a general reduction in the derived apparent $HD$ temperature from type 1 to type~2 sources, with both effects being due to extinction.

\section{Conclusions}\label{conc}

We present a spatially resolved stellar population study of the inner few tens of pc in a volume limited complete sample of nearby AGNs. This study was performed with adaptive optics assisted observations in the NIR. Our main results are summarised as follows:

\begin{itemize}
    
    \item In general, the galaxies display a very complex (e.g. many components are required to fit the underlying continuum) star formation history in the central region. 
    
    \item In terms of stellar populations, the light in the FoV of the galaxies is due to a considerable fraction of young ($xy$+$xyo$) stellar populations. The central region is dominated by an intermediate age population ($xiy$+$xio$), with a significant fraction of old bulge stars. 

    \item  The inner region of the galaxies has centrally peaked hot dust emission inside the unresolved region. On the other hand, for the $FC$ contribution we found that for nearly half of the sources this component reaches the highest values outside the unresolved nucleus. We have interpreted this as most likely due to dust emission rather than scattered light from the accretion disc.
    
    \item We found a correlation between the X-ray luminosity and $HD$, $FC$ and $A_V$. We attributed this correlation to the link they all have to the mass accretion rate onto the AGN: the AGN featureless continuum is produced by the accretion of gas, which is associated with more reddening; and it is also the source of heating for the circumnuclear dust, and thus increases the $HD$ contribution.
    
    \item We also found an anti-correlation of $HD$, $FC$, $A_V$, and $log(LBol_{obs})$ with $xiy$. We attribute it as a secondary effect, resulting from the fact that all of them are correlated, meaning that, when all their contributions increase, the result is a decreased contribution of the stellar population.
    
    \item A correlation of $log(LBol_{obs})$ with the mean age was also found.  We attribute this correlation as due to the gas that is ejected by the intermediate age population (at a 100\,pc scale) during the stellar evolution phases being used to feed the AGN.
    
    \item We also found that $HD$  and $FC$ components decrease outwards, $xy$ and $xyo$ increase from small to lager scales, while $xiy$, $xio$, and $xo$ are nearly constant over the analysed region.
    
\end{itemize}

In general, our results show that there is a significant fraction of young stellar populations in the inner region of the active galaxies, suggesting that the inner region of these sources is facing a rejuvenation process with the AGN impacting on the recent star formation, in the sense that it can be associated with the decrease of young star formation towards the nuclear region.

\section*{Acknowledgments}
We thank an anonymous referee for useful suggestions which helped to improve the paper. 
RR thanks to Conselho Nacional de Desenvolvimento Cient\'{i}fico e Tecnol\'ogico  ( CNPq, Proj. 311223/2020-6,  304927/2017-1 and  400352/2016-8), Funda\c{c}\~ao de amparo \`{a} pesquisa do Rio Grande do Sul (FAPERGS, Proj. 16/2551-0000251-7 and 19/1750-2), Coordena\c{c}\~ao de Aperfei\c{c}oamento de Pessoal de N\'{i}vel Superior (CAPES, Proj. 0001).

N.Z.D. acknowledges partial support from FONDECYT through project 3190769. CR acknowledges support from the Fondecyt Iniciacion grant 11190831 and ANID BASAL project FB210003. R.A.R acknowledges support from CNPq and Fapergs.  M.B thanks to CAPES (Finance code 001).

Based on observations obtained at the Gemini Observatory, 
which is operated by the Association of Universities for Research in Astronomy, Inc., under a cooperative agreement with the 
NSF on behalf of the Gemini partnership: the National Science Foundation (United States), the Science and Technology 
Facilities Council (United Kingdom), the National Research Council (Canada), CONICYT (Chile), the Australian Research 
Council (Australia), Minist\'erio da Ci\^encia e Tecnologia (Brazil) and south-eastCYT (Argentina).  

This research has made use of the NASA/IPAC Extragalactic Database (NED) which is operated by the Jet Propulsion Laboratory, California Institute of Technology, under contract with the National Aeronautics and Space Administration. We acknowledge the usage of the HyperLeda database (http://leda.univ-lyon1.fr).

This research made use of Photutils, an Astropy package for
detection and photometry of astronomical sources \citep{Bradley+19}

\section*{Data Availability}

The NIFS data used in this paper are available in the Gemini Observatory Archive \footnote{https://archive.gemini.edu/searchform}. They can also be shared by the first author under reasonable request.

%%%%%%%%%%%%%%%%%%%%%%%%%%%%%%%%%%%%%%%%%%%%%%%%%%



\bibliographystyle{mnras}
\bibliography{RefsZotero.bib}

\begin{thebibliography}{}
\makeatletter
\relax
\def\mn@urlcharsother{\let\do\@makeother \do\$\do\&\do\#\do\^\do\_\do\%\do\~}
\def\mn@doi{\begingroup\mn@urlcharsother \@ifnextchar [ {\mn@doi@}
  {\mn@doi@[]}}
\def\mn@doi@[#1]#2{\def\@tempa{#1}\ifx\@tempa\@empty \href
  {http://dx.doi.org/#2} {doi:#2}\else \href {http://dx.doi.org/#2} {#1}\fi
  \endgroup}
\def\mn@eprint#1#2{\mn@eprint@#1:#2::\@nil}
\def\mn@eprint@arXiv#1{\href {http://arxiv.org/abs/#1} {{\tt arXiv:#1}}}
\def\mn@eprint@dblp#1{\href {http://dblp.uni-trier.de/rec/bibtex/#1.xml}
  {dblp:#1}}
\def\mn@eprint@#1:#2:#3:#4\@nil{\def\@tempa {#1}\def\@tempb {#2}\def\@tempc
  {#3}\ifx \@tempc \@empty \let \@tempc \@tempb \let \@tempb \@tempa \fi \ifx
  \@tempb \@empty \def\@tempb {arXiv}\fi \@ifundefined
  {mn@eprint@\@tempb}{\@tempb:\@tempc}{\expandafter \expandafter \csname
  mn@eprint@\@tempb\endcsname \expandafter{\@tempc}}}

\bibitem[\protect\citeauthoryear{Ajello, Alexander, Greiner, Madejski, Gehrels
  \& Burlon}{Ajello et~al.}{2012}]{Ajello+12}
Ajello M.,  Alexander D.~M.,  Greiner J.,  Madejski G.~M.,  Gehrels N.,
  Burlon D.,  2012, \mn@doi [The Astrophysical Journal]
  {10.1088/0004-637X/749/1/21}, 749, 21

\bibitem[\protect\citeauthoryear{Alexander \& Hickox}{Alexander \&
  Hickox}{2012}]{Alexander+12}
Alexander D.~M.,  Hickox R.~C.,  2012, \mn@doi [New Astronomy Reviews]
  {10.1016/j.newar.2011.11.003}, 56, 93

\bibitem[\protect\citeauthoryear{{Alonso-Herrero} et~al.,}{{Alonso-Herrero}
  et~al.}{2019}]{Alonso-Herrero+19}
{Alonso-Herrero} A.,  et~al., 2019, \mn@doi [Astronomy \&amp; Astrophysics,
  Volume 628, id.A65, {$<$}NUMPAGES{$>$}17{$<$}/NUMPAGES{$>$} pp.]
  {10.1051/0004-6361/201935431}, 628, A65

\bibitem[\protect\citeauthoryear{{Alonso-Herrero} et~al.,}{{Alonso-Herrero}
  et~al.}{2020}]{Alonso-Herrero+20}
{Alonso-Herrero} A.,  et~al., 2020, \mn@doi [Astronomy \&amp; Astrophysics,
  Volume 639, id.A43, {$<$}NUMPAGES{$>$}17{$<$}/NUMPAGES{$>$} pp.]
  {10.1051/0004-6361/202037642}, 639, A43

\bibitem[\protect\citeauthoryear{Antonucci}{Antonucci}{1993}]{Antonucci+93}
Antonucci R.,  1993, \mn@doi [Annual Review of Astron and Astrophys]
  {10.1146/annurev.aa.31.090193.002353}, \href
  {http://adsabs.harvard.edu/abs/1993ARA\%26A..31..473A} {31, 473}

\bibitem[\protect\citeauthoryear{Asari, Cid~Fernandes, Stasi{\'n}ska,
  {Torres-Papaqui}, Mateus, Sodr{\'e}, Schoenell  \& Gomes}{Asari
  et~al.}{2007}]{Asari+07}
Asari N.~V.,  Cid~Fernandes R.,  Stasi{\'n}ska G.,  {Torres-Papaqui} J.~P.,
  Mateus A.,  Sodr{\'e} L.,  Schoenell W.,   Gomes J.~M.,  2007, \mn@doi
  [Monthly Notices of the Royal Astronomical Society]
  {10.1111/j.1365-2966.2007.12255.x}, 381, 263

\bibitem[\protect\citeauthoryear{{Astropy Collaboration} et~al.,}{{Astropy
  Collaboration} et~al.}{2018}]{AstropyCollaboration+18}
{Astropy Collaboration} et~al., 2018, \mn@doi [The Astronomical Journal]
  {10.3847/1538-3881/aabc4f}, 156, 123

\bibitem[\protect\citeauthoryear{Barbosa, {Storchi-Bergmann}, Cid~Fernandes,
  Winge  \& Schmitt}{Barbosa et~al.}{2009}]{Barbosa+09}
Barbosa F. K.~B.,  {Storchi-Bergmann} T.,  Cid~Fernandes R.,  Winge C.,
  Schmitt H.,  2009, \mn@doi [Monthly Notices of the Royal Astronomical
  Society] {10.1111/j.1365-2966.2009.14485.x}, 396, 2

\bibitem[\protect\citeauthoryear{Barbosa, {Storchi-Bergmann}, McGregor, Vale
  \& Rogemar~Riffel}{Barbosa et~al.}{2014}]{Barbosa+14}
Barbosa F. K.~B.,  {Storchi-Bergmann} T.,  McGregor P.,  Vale T.~B.,
  Rogemar~Riffel A.,  2014, \mn@doi [Monthly Notices of the Royal Astronomical
  Society] {10.1093/mnras/stu1637}, 445, 2353

\bibitem[\protect\citeauthoryear{Barvainis}{Barvainis}{1987}]{Barvainis+87}
Barvainis R.,  1987, \mn@doi [The Astrophysical Journal] {10.1086/165571}, 320,
  537

\bibitem[\protect\citeauthoryear{Bewketu~Belete et~al.,}{Bewketu~Belete
  et~al.}{2021}]{BewketuBelete+21}
Bewketu~Belete A.,  et~al., 2021, \mn@doi [Astronomy \&amp; Astrophysics,
  Volume 654, id.A24, {$<$}NUMPAGES{$>$}13{$<$}/NUMPAGES{$>$} pp.]
  {10.1051/0004-6361/202140492}, 654, A24

\bibitem[\protect\citeauthoryear{Bianchin et~al.,}{Bianchin
  et~al.}{2022}]{Bianchin+22}
Bianchin M.,  et~al., 2022, \mn@doi [Monthly Notices of the Royal Astronomical
  Society] {10.1093/mnras/stab3468}, 510, 639

\bibitem[\protect\citeauthoryear{Bieri, Dubois, Silk, Mamon  \& Gaibler}{Bieri
  et~al.}{2016}]{Bieri+16}
Bieri R.,  Dubois Y.,  Silk J.,  Mamon G.~A.,   Gaibler V.,  2016, \mn@doi
  [Monthly Notices of the Royal Astronomical Society] {10.1093/mnras/stv2551},
  455, 4166

\bibitem[\protect\citeauthoryear{Bradley et~al.,}{Bradley
  et~al.}{2019}]{Bradley+19}
Bradley L.,  et~al., 2019, Astropy/Photutils: V0.6, Zenodo,
  \mn@doi{10.5281/zenodo.2533376}

\bibitem[\protect\citeauthoryear{Burtscher et~al.,}{Burtscher
  et~al.}{2015}]{Burtscher+15}
Burtscher L.,  et~al., 2015, \mn@doi [Astronomy \&amp; Astrophysics, Volume
  578, id.A47, {$<$}NUMPAGES{$>$}15{$<$}/NUMPAGES{$>$} pp.]
  {10.1051/0004-6361/201525817}, 578, A47

\bibitem[\protect\citeauthoryear{Burtscher et~al.,}{Burtscher
  et~al.}{2021}]{Burtscher+21}
Burtscher L.,  et~al., 2021, arXiv:2105.05309 [astro-ph]

\bibitem[\protect\citeauthoryear{Caglar et~al.,}{Caglar
  et~al.}{2020}]{Caglar+20}
Caglar T.,  et~al., 2020, \mn@doi [Astronomy and Astrophysics]
  {10.1051/0004-6361/201936321}, 634, A114

\bibitem[\protect\citeauthoryear{Cappellari}{Cappellari}{2017}]{Cappellari+17}
Cappellari M.,  2017, \mn@doi [Monthly Notices of the Royal Astronomical
  Society] {10.1093/mnras/stw3020}, 466, 798

\bibitem[\protect\citeauthoryear{Cardamone, Moran  \& Kay}{Cardamone
  et~al.}{2007}]{Cardamone+07}
Cardamone C.~N.,  Moran E.~C.,   Kay L.~E.,  2007, \mn@doi [The Astronomical
  Journal] {10.1086/520801}, 134, 1263

\bibitem[\protect\citeauthoryear{Cardelli, Clayton  \& Mathis}{Cardelli
  et~al.}{1989}]{Cardelli+89}
Cardelli J.~A.,  Clayton G.~C.,   Mathis J.~S.,  1989, \mn@doi [The
  Astrophysical Journal] {10.1086/167900}, 345, 245

\bibitem[\protect\citeauthoryear{Cid~Fernandes}{Cid~Fernandes}{2018}]{CidFernandes+18}
Cid~Fernandes R.,  2018, \mn@doi [Monthly Notices of the Royal Astronomical
  Society] {10.1093/mnras/sty2012}, 480, 4480

\bibitem[\protect\citeauthoryear{Cid~Fernandes \& Terlevich}{Cid~Fernandes \&
  Terlevich}{1995}]{CidFernandes+95}
Cid~Fernandes Jr. R.,  Terlevich R.,  1995, \mn@doi [Monthly Notices of the
  Royal Astronomical Society] {10.1093/mnras/272.2.423}, 272, 423

\bibitem[\protect\citeauthoryear{Cid~Fernandes, Gu, Melnick, Terlevich,
  Terlevich, Kunth, Rodrigues~Lacerda  \& Joguet}{Cid~Fernandes
  et~al.}{2004}]{CidFernandes+04}
Cid~Fernandes R.,  Gu Q.,  Melnick J.,  Terlevich E.,  Terlevich R.,  Kunth D.,
   Rodrigues~Lacerda R.,   Joguet B.,  2004, \mn@doi [Monthly Notices of the
  RAS] {10.1111/j.1365-2966.2004.08321.x}, \href
  {http://adsabs.harvard.edu/abs/2004MNRAS.355..273C} {355, 273}

\bibitem[\protect\citeauthoryear{Cid~Fernandes, Mateus, Sodr{\'e},
  Stasi{\'n}ska  \& Gomes}{Cid~Fernandes et~al.}{2005}]{CidFernandes+05}
Cid~Fernandes R.,  Mateus A.,  Sodr{\'e} L.,  Stasi{\'n}ska G.,   Gomes J.~M.,
  2005, \mn@doi [Monthly Notices of the RAS]
  {10.1111/j.1365-2966.2005.08752.x}, \href
  {http://adsabs.harvard.edu/abs/2005MNRAS.358..363C} {358, 363}

\bibitem[\protect\citeauthoryear{Conroy}{Conroy}{2013}]{Conroy+13}
Conroy C.,  2013, \mn@doi [Annual Review of Astron and Astrophys]
  {10.1146/annurev-astro-082812-141017}, \href
  {http://adsabs.harvard.edu/abs/2013ARA\%26A..51..393C} {51, 393}

\bibitem[\protect\citeauthoryear{Crain et~al.,}{Crain et~al.}{2015}]{Crain+15}
Crain R.~A.,  et~al., 2015, \mn@doi [Monthly Notices of The Royal Astronomical
  Society] {10.1093/mnras/stv725}, 450, 1937

\bibitem[\protect\citeauthoryear{Croton et~al.,}{Croton
  et~al.}{2006}]{Croton+06}
Croton D.~J.,  et~al., 2006, \mn@doi [Monthly Notices of The Royal Astronomical
  Society] {10.1111/j.1365-2966.2005.09675.x}, 365, 11

\bibitem[\protect\citeauthoryear{Cuadra, Nayakshin, Springel  \&
  Di~Matteo}{Cuadra et~al.}{2006}]{Cuadra+06}
Cuadra J.,  Nayakshin S.,  Springel V.,   Di~Matteo T.,  2006, \mn@doi [Monthly
  Notices of the Royal Astronomical Society]
  {10.1111/j.1365-2966.2005.09837.x}, 366, 358

\bibitem[\protect\citeauthoryear{{Dahmer-Hahn} et~al.,}{{Dahmer-Hahn}
  et~al.}{2019a}]{Dahmer-Hahn+19a}
{Dahmer-Hahn} L.~G.,  et~al., 2019a, \mn@doi [Monthly Notices of the RAS]
  {10.1093/mnras/sty3051}, \href
  {http://adsabs.harvard.edu/abs/2019MNRAS.482.5211D} {482, 5211}

\bibitem[\protect\citeauthoryear{{Dahmer-Hahn} et~al.,}{{Dahmer-Hahn}
  et~al.}{2019b}]{Dahmer-Hahn+19}
{Dahmer-Hahn} L.~G.,  et~al., 2019b, \mn@doi [Monthly Notices of the Royal
  Astronomical Society] {10.1093/mnras/stz2453}, 489, 5653

\bibitem[\protect\citeauthoryear{{Dahmer-Hahn} et~al.,}{{Dahmer-Hahn}
  et~al.}{2021}]{Dahmer-Hahn+21}
{Dahmer-Hahn} L.~G.,  et~al., 2021, Stellar Populations in Local {{AGNs}}:
  Evidence for Enhanced Star Formation in the Inner 100pc

\bibitem[\protect\citeauthoryear{Dametto et~al.,}{Dametto
  et~al.}{2019}]{Dametto+19a}
Dametto N.~Z.,  et~al., 2019, \mn@doi [Monthly Notices of the RAS]
  {10.1093/mnras/sty2996}, \href
  {http://adsabs.harvard.edu/abs/2019MNRAS.482.4437D} {482, 4437}

\bibitem[\protect\citeauthoryear{Davies, Sternberg, Lehnert  \&
  {Tacconi-Garman}}{Davies et~al.}{2005}]{Davies+05}
Davies R.~I.,  Sternberg A.,  Lehnert M.~D.,   {Tacconi-Garman} L.~E.,  2005,
  \mn@doi [The Astrophysical Journal] {10.1086/444495}, 633, 105

\bibitem[\protect\citeauthoryear{Davies, M{\"u}ller~S{\'a}nchez, Genzel,
  Tacconi, Hicks, Friedrich  \& Sternberg}{Davies et~al.}{2007}]{Davies+07a}
Davies R.~I.,  M{\"u}ller~S{\'a}nchez F.,  Genzel R.,  Tacconi L.~J.,  Hicks E.
  K.~S.,  Friedrich S.,   Sternberg A.,  2007, \mn@doi [Astrophysical Journal]
  {10.1086/523032}, \href {http://adsabs.harvard.edu/abs/2007ApJ...671.1388D}
  {671, 1388}

\bibitem[\protect\citeauthoryear{Davies et~al.,}{Davies
  et~al.}{2015}]{Davies+15a}
Davies R.~I.,  et~al., 2015, \mn@doi [Astrophysical Journal]
  {10.1088/0004-637X/806/1/127}, \href
  {http://adsabs.harvard.edu/abs/2015ApJ...806..127D} {806, 127}

\bibitem[\protect\citeauthoryear{Davies et~al.,}{Davies
  et~al.}{2017}]{Davies+17}
Davies R.~I.,  et~al., 2017, \mn@doi [Monthly Notices of the Royal Astronomical
  Society] {10.1093/mnras/stx045}, 466, 4917

\bibitem[\protect\citeauthoryear{Davison \& Hinkley}{Davison \&
  Hinkley}{1997}]{Davison+97}
Davison A.~C.,  Hinkley D.~V.,  1997, Bootstrap {{Methods}} and Their
  {{Application}},
  /core/books/bootstrap-methods-and-their-application/ED2FD043579F27952363566DC09CBD6A,
  \mn@doi{10.1017/CBO9780511802843}

\bibitem[\protect\citeauthoryear{Di~Matteo, Springel  \& Hernquist}{Di~Matteo
  et~al.}{2005}]{DiMatteo+05}
Di~Matteo T.,  Springel V.,   Hernquist L.,  2005, \mn@doi [Nature]
  {10.1038/nature03335}, 433, 604

\bibitem[\protect\citeauthoryear{Dias, Coelho, Barbuy, Kerber  \& Idiart}{Dias
  et~al.}{2010}]{Dias+10}
Dias B.,  Coelho P.,  Barbuy B.,  Kerber L.,   Idiart T.,  2010, \mn@doi
  [Astronomy and Astrophysics] {10.1051/0004-6361/200912894}, 520, A85

\bibitem[\protect\citeauthoryear{Diniz, Riffel, {Storchi-Bergmann}  \&
  Riffel}{Diniz et~al.}{2019}]{Diniz+19}
Diniz M.~R.,  Riffel R.~A.,  {Storchi-Bergmann} T.,   Riffel R.,  2019, \mn@doi
  [Monthly Notices of the Royal Astronomical Society] {10.1093/mnras/stz1329},
  487, 3958

\bibitem[\protect\citeauthoryear{Dottori, D{\'i}az, Carranza, L{\'i}pari  \&
  Santos}{Dottori et~al.}{2005}]{Dottori+05}
Dottori H.,  D{\'i}az R.~J.,  Carranza G.,  L{\'i}pari S.,   Santos Jr. J.,
  2005, \mn@doi [The Astrophysical Journal] {10.1086/432714}, 628, L85

\bibitem[\protect\citeauthoryear{{El-Badry}, Wetzel, Geha, Hopkins, Kere{\v s},
  Chan  \& {Faucher-Gigu{\`e}re}}{{El-Badry} et~al.}{2016}]{El-Badry+16}
{El-Badry} K.,  Wetzel A.,  Geha M.,  Hopkins P.~F.,  Kere{\v s} D.,  Chan
  T.~K.,   {Faucher-Gigu{\`e}re} C.-A.,  2016, \mn@doi [The Astrophysical
  Journal] {10.3847/0004-637X/820/2/131}, 820, 131

\bibitem[\protect\citeauthoryear{Ellison, Teimoorinia, Rosario  \&
  Mendel}{Ellison et~al.}{2016}]{Ellison+16}
Ellison S.~L.,  Teimoorinia H.,  Rosario D.~J.,   Mendel J.~T.,  2016, \mn@doi
  [Monthly Notices of the Royal Astronomical Society] {10.1093/mnrasl/slw012},
  458, L34

\bibitem[\protect\citeauthoryear{Ellison et~al.,}{Ellison
  et~al.}{2021}]{Ellison+21}
Ellison S.~L.,  et~al., 2021, \mn@doi [Monthly Notices of the Royal
  Astronomical Society] {10.1093/mnrasl/slab047}, 505, L46

\bibitem[\protect\citeauthoryear{Esquej et~al.,}{Esquej
  et~al.}{2014}]{Esquej+14a}
Esquej P.,  et~al., 2014, \mn@doi [Astrophysical Journal]
  {10.1088/0004-637X/780/1/86}, \href
  {http://adsabs.harvard.edu/abs/2014ApJ...780...86E} {780, 86}

\bibitem[\protect\citeauthoryear{Fabian}{Fabian}{2012}]{Fabian+12}
Fabian A.~C.,  2012, \mn@doi [Annual Review of Astronomy and Astrophysics]
  {10.1146/annurev-astro-081811-125521}, 50, 455

\bibitem[\protect\citeauthoryear{Ferrarese \& Merritt}{Ferrarese \&
  Merritt}{2000}]{Ferrarese+00}
Ferrarese L.,  Merritt D.,  2000, \mn@doi [The Astrophysical Journal]
  {10.1086/312838}, 539, L9

\bibitem[\protect\citeauthoryear{Gallagher, Maiolino, Belfiore, Drory, Riffel
  \& Riffel}{Gallagher et~al.}{2019}]{Gallagher+19}
Gallagher R.,  Maiolino R.,  Belfiore F.,  Drory N.,  Riffel R.,   Riffel
  R.~A.,  2019, \mn@doi [Monthly Notices of the Royal Astronomical Society]
  {10.1093/mnras/stz564}, 485, 3409

\bibitem[\protect\citeauthoryear{Gao et~al.,}{Gao et~al.}{2020}]{Gao+20}
Gao F.,  et~al., 2020, \mn@doi [Astronomy \&amp; Astrophysics, Volume 637,
  id.A94, {$<$}NUMPAGES{$>$}16{$<$}/NUMPAGES{$>$} pp.]
  {10.1051/0004-6361/201937178}, 637, A94

\bibitem[\protect\citeauthoryear{{Garc{\'i}a-Burillo}
  et~al.,}{{Garc{\'i}a-Burillo} et~al.}{2019}]{Garcia-Burillo+19}
{Garc{\'i}a-Burillo} S.,  et~al., 2019, \mn@doi [Astronomy and Astrophysics]
  {10.1051/0004-6361/201936606}, 632, A61

\bibitem[\protect\citeauthoryear{{Garc{\'i}a-Burillo}
  et~al.,}{{Garc{\'i}a-Burillo} et~al.}{2021}]{Garcia-Burillo+21}
{Garc{\'i}a-Burillo} S.,  et~al., 2021, \mn@doi [Astronomy and Astrophysics]
  {10.1051/0004-6361/202141075}, 652, A98

\bibitem[\protect\citeauthoryear{{Garcia-Rissmann}, Vega, Asari, Cid~Fernandes,
  Schmitt, Gonz{\'a}lez~Delgado  \& {Storchi-Bergmann}}{{Garcia-Rissmann}
  et~al.}{2005}]{Garcia-Rissmann+05}
{Garcia-Rissmann} A.,  Vega L.~R.,  Asari N.~V.,  Cid~Fernandes R.,  Schmitt
  H.,  Gonz{\'a}lez~Delgado R.~M.,   {Storchi-Bergmann} T.,  2005, \mn@doi
  [Monthly Notices of the Royal Astronomical Society]
  {10.1111/j.1365-2966.2005.08957.x}, 359, 765

\bibitem[\protect\citeauthoryear{Gaspar, D{\'i}az, Mast, D'Ambra, Ag{\"u}ero
  \& G{\"u}nthardt}{Gaspar et~al.}{2019}]{Gaspar+19}
Gaspar G.,  D{\'i}az R.~J.,  Mast D.,  D'Ambra A.,  Ag{\"u}ero M.~P.,
  G{\"u}nthardt G.,  2019, \mn@doi [The Astronomical Journal]
  {10.3847/1538-3881/aaf4b9}, 157, 44

\bibitem[\protect\citeauthoryear{Ge, Yan, Cappellari, Mao, Li  \& Lu}{Ge
  et~al.}{2018}]{Ge+18}
Ge J.,  Yan R.,  Cappellari M.,  Mao S.,  Li H.,   Lu Y.,  2018, \mn@doi
  [Monthly Notices of the Royal Astronomical Society] {10.1093/mnras/sty1245},
  478, 2633

\bibitem[\protect\citeauthoryear{Gebhardt et~al.,}{Gebhardt
  et~al.}{2000}]{Gebhardt+00}
Gebhardt K.,  et~al., 2000, \mn@doi [The Astrophysical Journal]
  {10.1086/312840}, 539, L13

\bibitem[\protect\citeauthoryear{Goddard et~al.,}{Goddard
  et~al.}{2017}]{Goddard+17}
Goddard D.,  et~al., 2017, \mn@doi [Monthly Notices of the Royal Astronomical
  Society] {10.1093/mnras/stw3371}, 466, 4731

\bibitem[\protect\citeauthoryear{Gomes \& Papaderos}{Gomes \&
  Papaderos}{2017}]{Gomes+17}
Gomes J.~M.,  Papaderos P.,  2017, \mn@doi [Astronomy and Astrophysics]
  {10.1051/0004-6361/201628986}, 603, A63

\bibitem[\protect\citeauthoryear{Goulding et~al.,}{Goulding
  et~al.}{2018}]{Goulding+18}
Goulding A.~D.,  et~al., 2018, \mn@doi [Publications of the Astronomical
  Society of Japan] {10.1093/pasj/psx135}, 70

\bibitem[\protect\citeauthoryear{Granato \& Danese}{Granato \&
  Danese}{1994}]{Granato+94}
Granato G.~L.,  Danese L.,  1994, \mn@doi [Monthly Notices of the Royal
  Astronomical Society] {10.1093/mnras/268.1.235}, 268, 235

\bibitem[\protect\citeauthoryear{Granato, De~Zotti, Silva, Bressan  \&
  Danese}{Granato et~al.}{2004}]{Granato+04}
Granato G.~L.,  De~Zotti G.,  Silva L.,  Bressan A.,   Danese L.,  2004,
  \mn@doi [The Astrophysical Journal] {10.1086/379875}, 600, 580

\bibitem[\protect\citeauthoryear{Gratadour, Rouan, Mugnier, Fusco, Cl{\'e}net,
  Gendron  \& Lacombe}{Gratadour et~al.}{2006}]{Gratadour+06}
Gratadour D.,  Rouan D.,  Mugnier L.~M.,  Fusco T.,  Cl{\'e}net Y.,  Gendron
  E.,   Lacombe F.,  2006, \mn@doi [Astronomy and Astrophysics]
  {10.1051/0004-6361:20042191}, 446, 813

\bibitem[\protect\citeauthoryear{Greene \& Ho}{Greene \& Ho}{2006}]{Greene+06}
Greene J.~E.,  Ho L.~C.,  2006, \mn@doi [The Astrophysical Journal Letters]
  {10.1086/500507}, 641, L21

\bibitem[\protect\citeauthoryear{Harrison}{Harrison}{2017}]{Harrison+17}
Harrison C.~M.,  2017, \mn@doi [Nature Astronomy] {10.1038/s41550-017-0165}, 1,
  1

\bibitem[\protect\citeauthoryear{Heckman \& Best}{Heckman \&
  Best}{2014}]{Heckman+14}
Heckman T.~M.,  Best P.~N.,  2014, \mn@doi [Annual Review of Astronomy and
  Astrophysics, vol. 52, p.589-660] {10.1146/annurev-astro-081913-035722}, 52,
  589

\bibitem[\protect\citeauthoryear{Hennig, Riffel, Dors, Riffel,
  {Storchi-Bergmann}  \& Colina}{Hennig et~al.}{2018}]{Hennig+18}
Hennig M.~G.,  Riffel R.~A.,  Dors O.~L.,  Riffel R.,  {Storchi-Bergmann} T.,
  Colina L.,  2018, \mn@doi [Monthly Notices of the Royal Astronomical Society]
  {10.1093/mnras/sty547}, 477, 1086

\bibitem[\protect\citeauthoryear{Hickox, Mullaney, Alexander, Chen, Civano,
  Goulding  \& Hainline}{Hickox et~al.}{2014}]{Hickox+14}
Hickox R.~C.,  Mullaney J.~R.,  Alexander D.~M.,  Chen C.-T.~J.,  Civano F.~M.,
   Goulding A.~D.,   Hainline K.~N.,  2014, \mn@doi [The Astrophysical Journal]
  {10.1088/0004-637X/782/1/9}, 782, 9

\bibitem[\protect\citeauthoryear{Hinshaw et~al.,}{Hinshaw
  et~al.}{2013}]{Hinshaw+13}
Hinshaw G.,  et~al., 2013, \mn@doi [The Astrophysical Journal Supplement
  Series] {10.1088/0067-0049/208/2/19}, 208, 19

\bibitem[\protect\citeauthoryear{Ho, Greene, Filippenko  \& Sargent}{Ho
  et~al.}{2009}]{Ho+09}
Ho L.~C.,  Greene J.~E.,  Filippenko A.~V.,   Sargent W. L.~W.,  2009, \mn@doi
  [The Astrophysical Journal Supplement Series] {10.1088/0067-0049/183/1/1},
  183, 1

\bibitem[\protect\citeauthoryear{Hopkins}{Hopkins}{2012}]{Hopkins+12}
Hopkins P.~F.,  2012, \mn@doi [Monthly Notices of the RAS]
  {10.1111/j.1745-3933.2011.01179.x}, \href
  {http://adsabs.harvard.edu/abs/2012MNRAS.420L...8H} {420, L8}

\bibitem[\protect\citeauthoryear{Hopkins \& Elvis}{Hopkins \&
  Elvis}{2010}]{Hopkins+10a}
Hopkins P.~F.,  Elvis M.,  2010, \mn@doi [Monthly Notices of the Royal
  Astronomical Society] {10.1111/j.1365-2966.2009.15643.x}, 401, 7

\bibitem[\protect\citeauthoryear{Ichikawa, Ricci, Ueda, Matsuoka, Toba,
  Kawamuro, Trakhtenbrot  \& Koss}{Ichikawa et~al.}{2017}]{Ichikawa+17}
Ichikawa K.,  Ricci C.,  Ueda Y.,  Matsuoka K.,  Toba Y.,  Kawamuro T.,
  Trakhtenbrot B.,   Koss M.~J.,  2017, \mn@doi [The Astrophysical Journal]
  {10.3847/1538-4357/835/1/74}, 835, 74

\bibitem[\protect\citeauthoryear{Ishibashi \& Fabian}{Ishibashi \&
  Fabian}{2012}]{Ishibashi+12}
Ishibashi W.,  Fabian A.~C.,  2012, \mn@doi [Monthly Notices of the Royal
  Astronomical Society] {10.1111/j.1365-2966.2012.22074.x}, 427, 2998

\bibitem[\protect\citeauthoryear{Johnson, Leja, Conroy  \& Speagle}{Johnson
  et~al.}{2021}]{Johnson+21}
Johnson B.~D.,  Leja J.,  Conroy C.,   Speagle J.~S.,  2021, \mn@doi [The
  Astrophysical Journal Supplement Series] {10.3847/1538-4365/abef67}, 254, 22

\bibitem[\protect\citeauthoryear{Kawakatu \& Wada}{Kawakatu \&
  Wada}{2008}]{Kawakatu+08a}
Kawakatu N.,  Wada K.,  2008, \mn@doi [Astrophysical Journal] {10.1086/588574},
  \href {http://adsabs.harvard.edu/abs/2008ApJ...681...73K} {681, 73}

\bibitem[\protect\citeauthoryear{Kennicutt \& Evans}{Kennicutt \&
  Evans}{2012}]{Kennicutt+12}
Kennicutt R.~C.,  Evans N.~J.,  2012, \mn@doi [Annual Review of Astronomy and
  Astrophysics] {10.1146/annurev-astro-081811-125610}, 50, 531

\bibitem[\protect\citeauthoryear{King \& Pounds}{King \&
  Pounds}{2015}]{King+15}
King A.,  Pounds K.,  2015, \mn@doi [Annual Review of Astronomy and
  Astrophysics] {10.1146/annurev-astro-082214-122316}, 53, 115

\bibitem[\protect\citeauthoryear{Knapen, Shlosman  \& Peletier}{Knapen
  et~al.}{2000}]{Knapen+00}
Knapen J.~H.,  Shlosman I.,   Peletier R.~F.,  2000, \mn@doi [The Astrophysical
  Journal] {10.1086/308266}, 529, 93

\bibitem[\protect\citeauthoryear{Koleva, Prugniel, Ocvirk, Le~Borgne  \&
  Soubiran}{Koleva et~al.}{2008}]{Koleva+08}
Koleva M.,  Prugniel P.,  Ocvirk P.,  Le~Borgne D.,   Soubiran C.,  2008,
  \mn@doi [Monthly Notices of the Royal Astronomical Society]
  {10.1111/j.1365-2966.2008.12908.x}, 385, 1998

\bibitem[\protect\citeauthoryear{Koleva, Prugniel, Bouchard  \& Wu}{Koleva
  et~al.}{2009}]{Koleva+09}
Koleva M.,  Prugniel P.,  Bouchard A.,   Wu Y.,  2009, \mn@doi [Astronomy and
  Astrophysics] {10.1051/0004-6361/200811467}, 501, 1269

\bibitem[\protect\citeauthoryear{Kormendy \& Ho}{Kormendy \&
  Ho}{2013}]{Kormendy+13a}
Kormendy J.,  Ho L.~C.,  2013, \mn@doi [Annual Review of Astron and Astrophys]
  {10.1146/annurev-astro-082708-101811}, \href
  {http://adsabs.harvard.edu/abs/2013ARA\%26A..51..511K} {51, 511}

\bibitem[\protect\citeauthoryear{Koski}{Koski}{1978}]{Koski+78}
Koski A.~T.,  1978, \mn@doi [The Astrophysical Journal] {10.1086/156235}, 223,
  56

\bibitem[\protect\citeauthoryear{Koss et~al.,}{Koss et~al.}{2017}]{Koss+17}
Koss M.,  et~al., 2017, \mn@doi [The Astrophysical Journal]
  {10.3847/1538-4357/aa8ec9}, 850, 74

\bibitem[\protect\citeauthoryear{LaMassa, Heckman, Ptak, Martins, Wild,
  Sonnentrucker  \& Hornschemeier}{LaMassa et~al.}{2011}]{LaMassa+11}
LaMassa S.~M.,  Heckman T.~M.,  Ptak A.,  Martins L.,  Wild V.,  Sonnentrucker
  P.,   Hornschemeier A.,  2011, \mn@doi [The Astrophysical Journal]
  {10.1088/0004-637X/729/1/52}, 729, 52

\bibitem[\protect\citeauthoryear{Lopes, {Storchi-Bergmann}, Saraiva  \&
  Martini}{Lopes et~al.}{2007}]{Lopes+07}
Lopes R. D.~S.,  {Storchi-Bergmann} T.,  Saraiva M. d.~F.,   Martini P.,  2007,
  \mn@doi [The Astrophysical Journal] {10.1086/510064}, 655, 718

\bibitem[\protect\citeauthoryear{Maccagni, Morganti, Oosterloo  \&
  Mahony}{Maccagni et~al.}{2014}]{Maccagni+14}
Maccagni F.~M.,  Morganti R.,  Oosterloo T.~A.,   Mahony E.~K.,  2014, \mn@doi
  [Astronomy \&amp; Astrophysics, Volume 571, id.A67,
  {$<$}NUMPAGES{$>$}8{$<$}/NUMPAGES{$>$} pp.] {10.1051/0004-6361/201424334},
  571, A67

\bibitem[\protect\citeauthoryear{Madau \& Dickinson}{Madau \&
  Dickinson}{2014}]{Madau+14}
Madau P.,  Dickinson M.,  2014, \mn@doi [Annual Review of Astronomy and
  Astrophysics, vol. 52, p.415-486] {10.1146/annurev-astro-081811-125615}, 52,
  415

\bibitem[\protect\citeauthoryear{Magorrian et~al.,}{Magorrian
  et~al.}{1998}]{Magorrian+98}
Magorrian J.,  et~al., 1998, \mn@doi [The Astronomical Journal]
  {10.1086/300353}, 115, 2285

\bibitem[\protect\citeauthoryear{Maiolino et~al.,}{Maiolino
  et~al.}{2017}]{Maiolino+17}
Maiolino R.,  et~al., 2017, \mn@doi [Nature] {10.1038/nature21677}, 544, 202

\bibitem[\protect\citeauthoryear{Mallmann et~al.,}{Mallmann
  et~al.}{2018}]{Mallmann+18}
Mallmann N.~D.,  et~al., 2018, \mn@doi [Monthly Notices of the Royal
  Astronomical Society] {10.1093/mnras/sty1364}, 478, 5491

\bibitem[\protect\citeauthoryear{Maraston}{Maraston}{2005}]{Maraston+05}
Maraston C.,  2005, \mn@doi [Monthly Notices of the RAS]
  {10.1111/j.1365-2966.2005.09270.x}, \href
  {http://adsabs.harvard.edu/abs/2005MNRAS.362..799M} {362, 799}

\bibitem[\protect\citeauthoryear{Marian et~al.,}{Marian
  et~al.}{2020}]{Marian+20}
Marian V.,  et~al., 2020, \mn@doi [The Astrophysical Journal]
  {10.3847/1538-4357/abbd3e}, 904, 79

\bibitem[\protect\citeauthoryear{Marshall, Shabala, Krause, Pimbblet, Croton
  \& Owers}{Marshall et~al.}{2018}]{Marshall+18}
Marshall M.~A.,  Shabala S.~S.,  Krause M. G.~H.,  Pimbblet K.~A.,  Croton
  D.~J.,   Owers M.~S.,  2018, \mn@doi [Monthly Notices of the Royal
  Astronomical Society] {10.1093/mnras/stx2996}, 474, 3615

\bibitem[\protect\citeauthoryear{Martins, Riffel, {Rodr{\'i}guez-Ardila},
  Gruenwald  \& {de Souza}}{Martins et~al.}{2010}]{Martins+10a}
Martins L.~P.,  Riffel R.,  {Rodr{\'i}guez-Ardila} A.,  Gruenwald R.,   {de
  Souza} R.,  2010, \mn@doi [Monthly Notices of the RAS]
  {10.1111/j.1365-2966.2010.16817.x}, \href
  {http://adsabs.harvard.edu/abs/2010MNRAS.406.2185M} {406, 2185}

\bibitem[\protect\citeauthoryear{Mason et~al.,}{Mason et~al.}{2015}]{Mason+15a}
Mason R.~E.,  et~al., 2015, \mn@doi [Astrophysical Journal, Supplement]
  {10.1088/0067-0049/217/1/13}, \href
  {http://adsabs.harvard.edu/abs/2015ApJS..217...13M} {217, 13}

\bibitem[\protect\citeauthoryear{Mauduit \& Mamon}{Mauduit \&
  Mamon}{2007}]{Mauduit+07}
Mauduit J.-C.,  Mamon G.~A.,  2007, \mn@doi [Astronomy and Astrophysics, Volume
  475, Issue 1, November III 2007, pp.169-185] {10.1051/0004-6361:20077721},
  475, 169

\bibitem[\protect\citeauthoryear{McGregor et~al.,}{McGregor
  et~al.}{2003}]{McGregor+03}
McGregor P.~J.,  et~al., 2003, \mn@doi [Instrument Design and Performance for
  Optical/Infrared Ground-based Telescopes. Edited by Iye, Masanori; Moorwood,
  Alan F. M. Proceedings of the SPIE] {10.1117/12.459448}, 4841, 1581

\bibitem[\protect\citeauthoryear{Nayakshin \& Zubovas}{Nayakshin \&
  Zubovas}{2012}]{Nayakshin+12}
Nayakshin S.,  Zubovas K.,  2012, \mn@doi [Monthly Notices of the Royal
  Astronomical Society] {10.1111/j.1365-2966.2012.21950.x}, 427, 372

\bibitem[\protect\citeauthoryear{Novak, Ostriker  \& Ciotti}{Novak
  et~al.}{2011}]{Novak+11}
Novak G.~S.,  Ostriker J.~P.,   Ciotti L.,  2011, \mn@doi [The Astrophysical
  Journal] {10.1088/0004-637X/737/1/26}, 737, 26

\bibitem[\protect\citeauthoryear{Ocvirk, Pichon, Lan{\c c}on  \&
  Thi{\'e}baut}{Ocvirk et~al.}{2006}]{Ocvirk+06}
Ocvirk P.,  Pichon C.,  Lan{\c c}on A.,   Thi{\'e}baut E.,  2006, \mn@doi
  [Monthly Notices of the Royal Astronomical Society]
  {10.1111/j.1365-2966.2005.09182.x}, 365, 46

\bibitem[\protect\citeauthoryear{Oh et~al.,}{Oh et~al.}{2018}]{Oh+18}
Oh K.,  et~al., 2018, \mn@doi [The Astrophysical Journal Supplement Series]
  {10.3847/1538-4365/aaa7fd}, 235, 4

\bibitem[\protect\citeauthoryear{Oliva, Origlia, Maiolino  \& Moorwood}{Oliva
  et~al.}{1999}]{Oliva+99}
Oliva E.,  Origlia L.,  Maiolino R.,   Moorwood A. F.~M.,  1999, Astronomy and
  Astrophysics, 350, 9

\bibitem[\protect\citeauthoryear{Owen}{Owen}{2007}]{Owen+07}
Owen A.,  2007, \mn@doi [Contemp. Math.] {10.1090/conm/443/08555}, 443

\bibitem[\protect\citeauthoryear{Pan et~al.,}{Pan et~al.}{2019}]{Pan+19}
Pan H.-A.,  et~al., 2019, \mn@doi [The Astrophysical Journal]
  {10.3847/1538-4357/ab311c}, 881, 119

\bibitem[\protect\citeauthoryear{Prieto, {Fernandez-Ontiveros}, Bruzual,
  Burkert, Schartmann  \& Charlot}{Prieto et~al.}{2019}]{Prieto+19}
Prieto M.~A.,  {Fernandez-Ontiveros} J.~A.,  Bruzual G.,  Burkert A.,
  Schartmann M.,   Charlot S.,  2019, \mn@doi [Monthly Notices of the Royal
  Astronomical Society] {10.1093/mnras/stz579}, 485, 3264

\bibitem[\protect\citeauthoryear{{Ramos-Almeida} \& Ricci}{{Ramos-Almeida} \&
  Ricci}{2017}]{Ramos-Almeida+17}
{Ramos-Almeida} C.~R.,  Ricci C.,  2017, \mn@doi [Nature Astronomy]
  {10.1038/s41550-017-0232-z}, 1, 679

\bibitem[\protect\citeauthoryear{Rees}{Rees}{1989}]{Rees+89}
Rees M.~J.,  1989, \mn@doi [Monthly Notices of the Royal Astronomical Society]
  {10.1093/mnras/239.1.1P}, 239, 1P

\bibitem[\protect\citeauthoryear{Reichard, Heckman, Rudnick, Brinchmann,
  Kauffmann  \& Wild}{Reichard et~al.}{2009}]{Reichard+09}
Reichard T.~A.,  Heckman T.~M.,  Rudnick G.,  Brinchmann J.,  Kauffmann G.,
  Wild V.,  2009, \mn@doi [The Astrophysical Journal]
  {10.1088/0004-637X/691/2/1005}, 691, 1005

\bibitem[\protect\citeauthoryear{Ricci, Ueda, Koss, Trakhtenbrot, Bauer  \&
  Gandhi}{Ricci et~al.}{2015}]{Ricci+15}
Ricci C.,  Ueda Y.,  Koss M.~J.,  Trakhtenbrot B.,  Bauer F.~E.,   Gandhi P.,
  2015, \mn@doi [The Astrophysical Journal] {10.1088/2041-8205/815/1/L13}, 815,
  L13

\bibitem[\protect\citeauthoryear{Ricci et~al.,}{Ricci et~al.}{2017}]{Ricci+17}
Ricci C.,  et~al., 2017, \mn@doi [The Astrophysical Journal Supplement Series]
  {10.3847/1538-4365/aa96ad}, 233, 17

\bibitem[\protect\citeauthoryear{Riffel, {Rodr{\'i}guez-Ardila}  \&
  Pastoriza}{Riffel et~al.}{2006}]{Riffel+06}
Riffel R.,  {Rodr{\'i}guez-Ardila} A.,   Pastoriza M.~G.,  2006, \mn@doi
  [Astronomy and Astrophysics] {10.1051/0004-6361:20065291}, \href
  {http://adsabs.harvard.edu/abs/2006A\%26A...457...61R} {457, 61}

\bibitem[\protect\citeauthoryear{Riffel, Pastoriza, {Rodr{\'i}guez-Ardila}  \&
  Maraston}{Riffel et~al.}{2007}]{Riffel+07}
Riffel R.,  Pastoriza M.~G.,  {Rodr{\'i}guez-Ardila} A.,   Maraston C.,  2007,
  \mn@doi [Astrophysical Journal, Letters] {10.1086/517999}, \href
  {http://adsabs.harvard.edu/abs/2007ApJ...659L.103R} {659, L103}

\bibitem[\protect\citeauthoryear{Riffel, Pastoriza, {Rodr{\'i}guez-Ardila}  \&
  Bonatto}{Riffel et~al.}{2009}]{Riffel+09}
Riffel R.,  Pastoriza M.~G.,  {Rodr{\'i}guez-Ardila} A.,   Bonatto C.,  2009,
  \mn@doi [Monthly Notices of the RAS] {10.1111/j.1365-2966.2009.15448.x},
  \href {http://adsabs.harvard.edu/abs/2009MNRAS.400..273R} {400, 273}

\bibitem[\protect\citeauthoryear{Riffel, {Storchi-Bergmann}, Riffel  \&
  Pastoriza}{Riffel et~al.}{2010}]{Riffel+10}
Riffel R.~A.,  {Storchi-Bergmann} T.,  Riffel R.,   Pastoriza M.~G.,  2010,
  \mn@doi [Astrophysical Journal] {10.1088/0004-637X/713/1/469}, \href
  {http://adsabs.harvard.edu/abs/2010ApJ...713..469R} {713, 469}

\bibitem[\protect\citeauthoryear{Riffel, {Ruschel-Dutra}, Pastoriza,
  {Rodr{\'i}guez-Ardila}, Santos, Bonatto  \& Ducati}{Riffel
  et~al.}{2011a}]{Riffel+11a}
Riffel R.,  {Ruschel-Dutra} D.,  Pastoriza M.~G.,  {Rodr{\'i}guez-Ardila} A.,
  Santos Jr. J. F.~C.,  Bonatto C.~J.,   Ducati J.~R.,  2011a, \mn@doi [Monthly
  Notices of the RAS] {10.1111/j.1365-2966.2010.17647.x}, \href
  {http://adsabs.harvard.edu/abs/2011MNRAS.410.2714R} {410, 2714}

\bibitem[\protect\citeauthoryear{Riffel, Bonatto, Cid~Fernandes, Pastoriza  \&
  Balbinot}{Riffel et~al.}{2011b}]{Riffel+11c}
Riffel R.,  Bonatto C.,  Cid~Fernandes R.,  Pastoriza M.~G.,   Balbinot E.,
  2011b, \mn@doi [Monthly Notices of the RAS]
  {10.1111/j.1365-2966.2010.17819.x}, \href
  {http://adsabs.harvard.edu/abs/2011MNRAS.411.1897R} {411, 1897}

\bibitem[\protect\citeauthoryear{Riffel, Riffel, Ferrari  \&
  {Storchi-Bergmann}}{Riffel et~al.}{2011c}]{Riffel+11}
Riffel R.,  Riffel R.~A.,  Ferrari F.,   {Storchi-Bergmann} T.,  2011c, \mn@doi
  [Monthly Notices of the RAS] {10.1111/j.1365-2966.2011.19061.x}, \href
  {http://adsabs.harvard.edu/abs/2011MNRAS.416..493R} {416, 493}

\bibitem[\protect\citeauthoryear{Riffel et~al.,}{Riffel
  et~al.}{2013a}]{Riffel+13}
Riffel R.~A.,  et~al., 2013a, \mn@doi [Monthly Notices of the RAS]
  {10.1093/mnras/sts536}, \href
  {http://adsabs.harvard.edu/abs/2013MNRAS.429.2587R} {429, 2587}

\bibitem[\protect\citeauthoryear{Riffel, {Rodr{\'i}guez-Ardila}, Aleman,
  Brotherton, Pastoriza, Bonatto  \& Dors}{Riffel et~al.}{2013b}]{Riffel+13a}
Riffel R.,  {Rodr{\'i}guez-Ardila} A.,  Aleman I.,  Brotherton M.~S.,
  Pastoriza M.~G.,  Bonatto C.,   Dors O.~L.,  2013b, \mn@doi [Monthly Notices
  of the RAS] {10.1093/mnras/stt026}, \href
  {http://adsabs.harvard.edu/abs/2013MNRAS.430.2002R} {430, 2002}

\bibitem[\protect\citeauthoryear{Riffel, Vale, {Storchi-Bergmann}  \&
  McGregor}{Riffel et~al.}{2014}]{Riffel+14}
Riffel R.~A.,  Vale T.~B.,  {Storchi-Bergmann} T.,   McGregor P.~J.,  2014,
  \mn@doi [Monthly Notices of the Royal Astronomical Society]
  {10.1093/mnras/stu843}, 442, 656

\bibitem[\protect\citeauthoryear{Riffel et~al.,}{Riffel
  et~al.}{2015}]{Riffel+15}
Riffel R.,  et~al., 2015, \mn@doi [Monthly Notices of the RAS]
  {10.1093/mnras/stv866}, \href
  {http://adsabs.harvard.edu/abs/2015MNRAS.450.3069R} {450, 3069}

\bibitem[\protect\citeauthoryear{Riffel et~al.,}{Riffel
  et~al.}{2016}]{Riffel+16a}
Riffel R.~A.,  et~al., 2016, \mn@doi [Monthly Notices of the Royal Astronomical
  Society] {10.1093/mnras/stw1609}, 461, 4192

\bibitem[\protect\citeauthoryear{Riffel, {Storchi-Bergmann}, Riffel,
  {Dahmer-Hahn}, Diniz, Sch{\"o}nell  \& Dametto}{Riffel
  et~al.}{2017}]{Riffel+17}
Riffel R.~A.,  {Storchi-Bergmann} T.,  Riffel R.,  {Dahmer-Hahn} L.~G.,  Diniz
  M.~R.,  Sch{\"o}nell A.~J.,   Dametto N.~Z.,  2017, \mn@doi [Monthly Notices
  of the Royal Astronomical Society] {10.1093/mnras/stx1308}, 470, 992

\bibitem[\protect\citeauthoryear{Riffel et~al.,}{Riffel
  et~al.}{2018}]{Riffel+18}
Riffel R.~A.,  et~al., 2018, \mn@doi [Monthly Notices of the Royal Astronomical
  Society] {10.1093/mnras/stx2857}, 474, 1373

\bibitem[\protect\citeauthoryear{Riffel et~al.,}{Riffel
  et~al.}{2019}]{Riffel+19}
Riffel R.,  et~al., 2019, \mn@doi [Monthly Notices of the Royal Astronomical
  Society] {10.1093/mnras/stz1077}, 486, 3228

\bibitem[\protect\citeauthoryear{Riffel et~al.,}{Riffel
  et~al.}{2021a}]{Riffel+21}
Riffel R.,  et~al., 2021a, \mn@doi [Monthly Notices of the Royal Astronomical
  Society] {10.1093/mnras/staa3907}, 501, 4064

\bibitem[\protect\citeauthoryear{Riffel et~al.,}{Riffel
  et~al.}{2021b}]{Riffel+21a}
Riffel R.~A.,  et~al., 2021b, \mn@doi [Monthly Notices of the Royal
  Astronomical Society] {10.1093/mnras/stab998}, 504, 3265

\bibitem[\protect\citeauthoryear{{Rodr{\'i}guez-Ardila}, Pastoriza, Viegas,
  Sigut  \& Pradhan}{{Rodr{\'i}guez-Ardila} et~al.}{2004}]{Rodriguez-Ardila+04}
{Rodr{\'i}guez-Ardila} A.,  Pastoriza M.~G.,  Viegas S.,  Sigut T. a.~A.,
  Pradhan A.~K.,  2004, \mn@doi [Astronomy and Astrophysics, v.425, p.457-474
  (2004)] {10.1051/0004-6361:20034285}, 425, 457

\bibitem[\protect\citeauthoryear{{Rodr{\'i}guez-Ardila}, Riffel  \&
  Pastoriza}{{Rodr{\'i}guez-Ardila} et~al.}{2005}]{Rodriguez-Ardila+05}
{Rodr{\'i}guez-Ardila} A.,  Riffel R.,   Pastoriza M.~G.,  2005, \mn@doi
  [Monthly Notices of the Royal Astronomical Society]
  {10.1111/j.1365-2966.2005.09638.x}, 364, 1041

\bibitem[\protect\citeauthoryear{{Ruschel-Dutra}, Rodr{\'i}guez~Espinosa,
  Gonz{\'a}lez~Mart{\'i}n, Pastoriza  \& Riffel}{{Ruschel-Dutra}
  et~al.}{2017}]{Ruschel-Dutra+17a}
{Ruschel-Dutra} D.,  Rodr{\'i}guez~Espinosa J.~M.,  Gonz{\'a}lez~Mart{\'i}n O.,
   Pastoriza M.,   Riffel R.,  2017, \mn@doi [Monthly Notices of the RAS]
  {10.1093/mnras/stw3276}, \href
  {http://adsabs.harvard.edu/abs/2017MNRAS.466.3353R} {466, 3353}

\bibitem[\protect\citeauthoryear{Rybicki \& Lightman}{Rybicki \&
  Lightman}{1979}]{Rybicki+79}
Rybicki G.~B.,  Lightman A.~P.,  1979, New York, Wiley-Interscience, 1979. 393
  p.

\bibitem[\protect\citeauthoryear{Salaris, Weiss, Cassar{\`a}, Piovan  \&
  Chiosi}{Salaris et~al.}{2014}]{Salaris+14a}
Salaris M.,  Weiss A.,  Cassar{\`a} L.~P.,  Piovan L.,   Chiosi C.,  2014,
  \mn@doi [Astronomy and Astrophysics] {10.1051/0004-6361/201423542}, \href
  {http://adsabs.harvard.edu/abs/2014A\%26A...565A...9S} {565, A9}

\bibitem[\protect\citeauthoryear{S{\'a}nchez et~al.,}{S{\'a}nchez
  et~al.}{2016}]{Sanchez+16}
S{\'a}nchez S.~F.,  et~al., 2016, Revista Mexicana de Astronomia y Astrofisica,
  52, 21

\bibitem[\protect\citeauthoryear{Sarzi, Allard, Knapen  \& Mazzuca}{Sarzi
  et~al.}{2007}]{Sarzi+07a}
Sarzi M.,  Allard E.~L.,  Knapen J.~H.,   Mazzuca L.~M.,  2007, \mn@doi
  [Monthly Notices of the RAS] {10.1111/j.1365-2966.2007.12177.x}, \href
  {http://adsabs.harvard.edu/abs/2007MNRAS.380..949S} {380, 949}

\bibitem[\protect\citeauthoryear{Schawinski, Koss, Berney  \&
  Sartori}{Schawinski et~al.}{2015}]{Schawinski+15}
Schawinski K.,  Koss M.,  Berney S.,   Sartori L.~F.,  2015, \mn@doi [Monthly
  Notices of the Royal Astronomical Society] {10.1093/mnras/stv1136}, 451, 2517

\bibitem[\protect\citeauthoryear{Schaye et~al.,}{Schaye
  et~al.}{2015}]{Schaye+15}
Schaye J.,  et~al., 2015, \mn@doi [Monthly Notices of the Royal Astronomical
  Society] {10.1093/mnras/stu2058}, 446, 521

\bibitem[\protect\citeauthoryear{Schlegel, Finkbeiner  \& Davis}{Schlegel
  et~al.}{1998}]{Schlegel+98}
Schlegel D.~J.,  Finkbeiner D.~P.,   Davis M.,  1998, \mn@doi [The
  Astrophysical Journal] {10.1086/305772}, 500, 525

\bibitem[\protect\citeauthoryear{Sch{\"o}nell, {Storchi-Bergmann}, Riffel  \&
  Riffel}{Sch{\"o}nell et~al.}{2017}]{Schonell+17}
Sch{\"o}nell Jr. A.~J.,  {Storchi-Bergmann} T.,  Riffel R.~A.,   Riffel R.,
  2017, \mn@doi [Monthly Notices of the Royal Astronomical Society]
  {10.1093/mnras/stw2263}, 464, 1771

\bibitem[\protect\citeauthoryear{Springel et~al.,}{Springel
  et~al.}{2005}]{Springel+05}
Springel V.,  et~al., 2005, \mn@doi [Nature] {10.1038/nature03597}, 435, 629

\bibitem[\protect\citeauthoryear{{Storchi-Bergmann} \&
  {Schnorr-M{\"u}ller}}{{Storchi-Bergmann} \&
  {Schnorr-M{\"u}ller}}{2019}]{Storchi-Bergmann+19}
{Storchi-Bergmann} T.,  {Schnorr-M{\"u}ller} A.,  2019, \mn@doi [Nature
  Astronomy] {10.1038/s41550-018-0611-0}, 3, 48

\bibitem[\protect\citeauthoryear{{Storchi-Bergmann}, Gonz{\'a}lez~Delgado,
  Schmitt, Cid~Fernandes  \& Heckman}{{Storchi-Bergmann}
  et~al.}{2001}]{Storchi-Bergmann+01a}
{Storchi-Bergmann} T.,  Gonz{\'a}lez~Delgado R.~M.,  Schmitt H.~R.,
  Cid~Fernandes R.,   Heckman T.,  2001, \mn@doi [Astrophysical Journal]
  {10.1086/322290}, \href {http://adsabs.harvard.edu/abs/2001ApJ...559..147S}
  {559, 147}

\bibitem[\protect\citeauthoryear{{Storchi-Bergmann}, McGregor, Riffel,
  Sim{\~o}es~Lopes, Beck  \& Dopita}{{Storchi-Bergmann}
  et~al.}{2009}]{Storchi-Bergmann+09a}
{Storchi-Bergmann} T.,  McGregor P.~J.,  Riffel R.~A.,  Sim{\~o}es~Lopes R.,
  Beck T.,   Dopita M.,  2009, \mn@doi [Monthly Notices of the RAS]
  {10.1111/j.1365-2966.2009.14388.x}, \href
  {http://adsabs.harvard.edu/abs/2009MNRAS.394.1148S} {394, 1148}

\bibitem[\protect\citeauthoryear{{Storchi-Bergmann}, Lopes, McGregor, Riffel,
  Beck  \& Martini}{{Storchi-Bergmann} et~al.}{2010}]{Storchi-Bergmann+10a}
{Storchi-Bergmann} T.,  Lopes R. D.~S.,  McGregor P.~J.,  Riffel R.~A.,  Beck
  T.,   Martini P.,  2010, \mn@doi [Monthly Notices of the RAS]
  {10.1111/j.1365-2966.2009.15962.x}, \href
  {http://adsabs.harvard.edu/abs/2010MNRAS.402..819S} {402, 819}

\bibitem[\protect\citeauthoryear{{Storchi-Bergmann}, Riffel, Riffel, Diniz,
  Borges~Vale  \& McGregor}{{Storchi-Bergmann}
  et~al.}{2012}]{Storchi-Bergmann+12a}
{Storchi-Bergmann} T.,  Riffel R.~A.,  Riffel R.,  Diniz M.~R.,  Borges~Vale
  T.,   McGregor P.~J.,  2012, \mn@doi [Astrophysical Journal]
  {10.1088/0004-637X/755/2/87}, \href
  {http://adsabs.harvard.edu/abs/2012ApJ...755...87S} {755, 87}

\bibitem[\protect\citeauthoryear{Terlevich, Diaz  \& Terlevich}{Terlevich
  et~al.}{1990}]{Terlevich+90a}
Terlevich E.,  Diaz A.~I.,   Terlevich R.,  1990, \mn@doi [Monthly Notices of
  the RAS] {10.1093/mnras/242.3.271}, \href
  {http://adsabs.harvard.edu/abs/1990MNRAS.242..271T} {242, 271}

\bibitem[\protect\citeauthoryear{Terrazas et~al.,}{Terrazas
  et~al.}{2020}]{Terrazas+20}
Terrazas B.~A.,  et~al., 2020, \mn@doi [Monthly Notices of the Royal
  Astronomical Society] {10.1093/mnras/staa374}, 493, 1888

\bibitem[\protect\citeauthoryear{Tody}{Tody}{1986}]{Tody+86}
Tody D.,  1986, \mn@doi [IN: Instrumentation in astronomy VI; Proceedings of
  the Meeting, Tucson, AZ, Mar. 4-8, 1986. Part 2 (A87-36376 15-35).
  Bellingham, WA, Society of Photo-Optical Instrumentation Engineers]
  {10.1117/12.968154}, 627, 733

\bibitem[\protect\citeauthoryear{Tody}{Tody}{1993}]{Tody+93}
Tody D.,  1993, Astronomical Data Analysis Software and Systems II, A.S.P.
  Conference Series,, R. J. Hanisch, R. J. V. Brissenden, and Jeannette Barnes,
  eds.,, 52, 173

\bibitem[\protect\citeauthoryear{Tojeiro, Heavens, Jimenez  \& Panter}{Tojeiro
  et~al.}{2007}]{Tojeiro+07}
Tojeiro R.,  Heavens A.~F.,  Jimenez R.,   Panter B.,  2007, \mn@doi [Monthly
  Notices of the Royal Astronomical Society]
  {10.1111/j.1365-2966.2007.12323.x}, 381, 1252

\bibitem[\protect\citeauthoryear{Trussler, Maiolino, Maraston, Peng, Thomas,
  Goddard  \& Lian}{Trussler et~al.}{2020}]{Trussler+20}
Trussler J.,  Maiolino R.,  Maraston C.,  Peng Y.,  Thomas D.,  Goddard D.,
  Lian J.,  2020, \mn@doi [Monthly Notices of the Royal Astronomical Society]
  {10.1093/mnras/stz3286}, 491, 5406

\bibitem[\protect\citeauthoryear{Villforth et~al.,}{Villforth
  et~al.}{2014}]{Villforth+14}
Villforth C.,  et~al., 2014, \mn@doi [Monthly Notices of the Royal Astronomical
  Society] {10.1093/mnras/stu173}, 439, 3342

\bibitem[\protect\citeauthoryear{Vogelsberger et~al.,}{Vogelsberger
  et~al.}{2014}]{Vogelsberger+14}
Vogelsberger M.,  et~al., 2014, \mn@doi [Nature] {10.1038/nature13316}, 509,
  177

\bibitem[\protect\citeauthoryear{Walcher, Groves, Budav{\'a}ri  \&
  Dale}{Walcher et~al.}{2011}]{Walcher+11}
Walcher J.,  Groves B.,  Budav{\'a}ri T.,   Dale D.,  2011, \mn@doi
  [Astrophysics and Space Science] {10.1007/s10509-010-0458-z}, 331, 1

\bibitem[\protect\citeauthoryear{Wang \& Loeb}{Wang \& Loeb}{2018}]{Wang+18}
Wang X.,  Loeb A.,  2018, \mn@doi [New Astronomy]
  {10.1016/j.newast.2017.12.004}, 61, 95

\bibitem[\protect\citeauthoryear{Waskom}{Waskom}{2021}]{Waskom+21}
Waskom M.~L.,  2021, \mn@doi [Journal of Open Source Software]
  {10.21105/joss.03021}, 6, 3021

\bibitem[\protect\citeauthoryear{Wilkinson, Maraston, Goddard, Thomas  \&
  Parikh}{Wilkinson et~al.}{2017}]{Wilkinson+17}
Wilkinson D.~M.,  Maraston C.,  Goddard D.,  Thomas D.,   Parikh T.,  2017,
  \mn@doi [Monthly Notices of the Royal Astronomical Society]
  {10.1093/mnras/stx2215}, 472, 4297

\bibitem[\protect\citeauthoryear{Worthey, Faber, Gonzalez  \& Burstein}{Worthey
  et~al.}{1994}]{Worthey+94a}
Worthey G.,  Faber S.~M.,  Gonzalez J.~J.,   Burstein D.,  1994, \mn@doi
  [Astrophysical Journal, Supplement] {10.1086/192087}, \href
  {http://adsabs.harvard.edu/abs/1994ApJS...94..687W} {94, 687}

\bibitem[\protect\citeauthoryear{Yang, Xie, Yuan, Zdziarski, Gierli{\'n}ski, Ho
   \& Yu}{Yang et~al.}{2015}]{Yang+15}
Yang Q.-X.,  Xie F.-G.,  Yuan F.,  Zdziarski A.~A.,  Gierli{\'n}ski M.,  Ho
  L.~C.,   Yu Z.,  2015, \mn@doi [Monthly Notices of the Royal Astronomical
  Society] {10.1093/mnras/stu2571}, 447, 1692

\bibitem[\protect\citeauthoryear{Zubovas \& Bourne}{Zubovas \&
  Bourne}{2017}]{Zubovas+17a}
Zubovas K.,  Bourne M.~A.,  2017, \mn@doi [Monthly Notices of the RAS]
  {10.1093/mnras/stx787}, \href
  {http://adsabs.harvard.edu/abs/2017arXiv170310782Z} {468, 4956}

\bibitem[\protect\citeauthoryear{Zubovas, Nayakshin, King  \&
  Wilkinson}{Zubovas et~al.}{2013}]{Zubovas+13}
Zubovas K.,  Nayakshin S.,  King A.,   Wilkinson M.,  2013, \mn@doi [Monthly
  Notices of the Royal Astronomical Society] {10.1093/mnras/stt952}, 433, 3079

\makeatother
\end{thebibliography}

%\appendix 
%\include{currentprofiles}
%\include{previousProfiles}

\bsp	% typesetting comment
\label{lastpage}

\includepdf[pages=-]{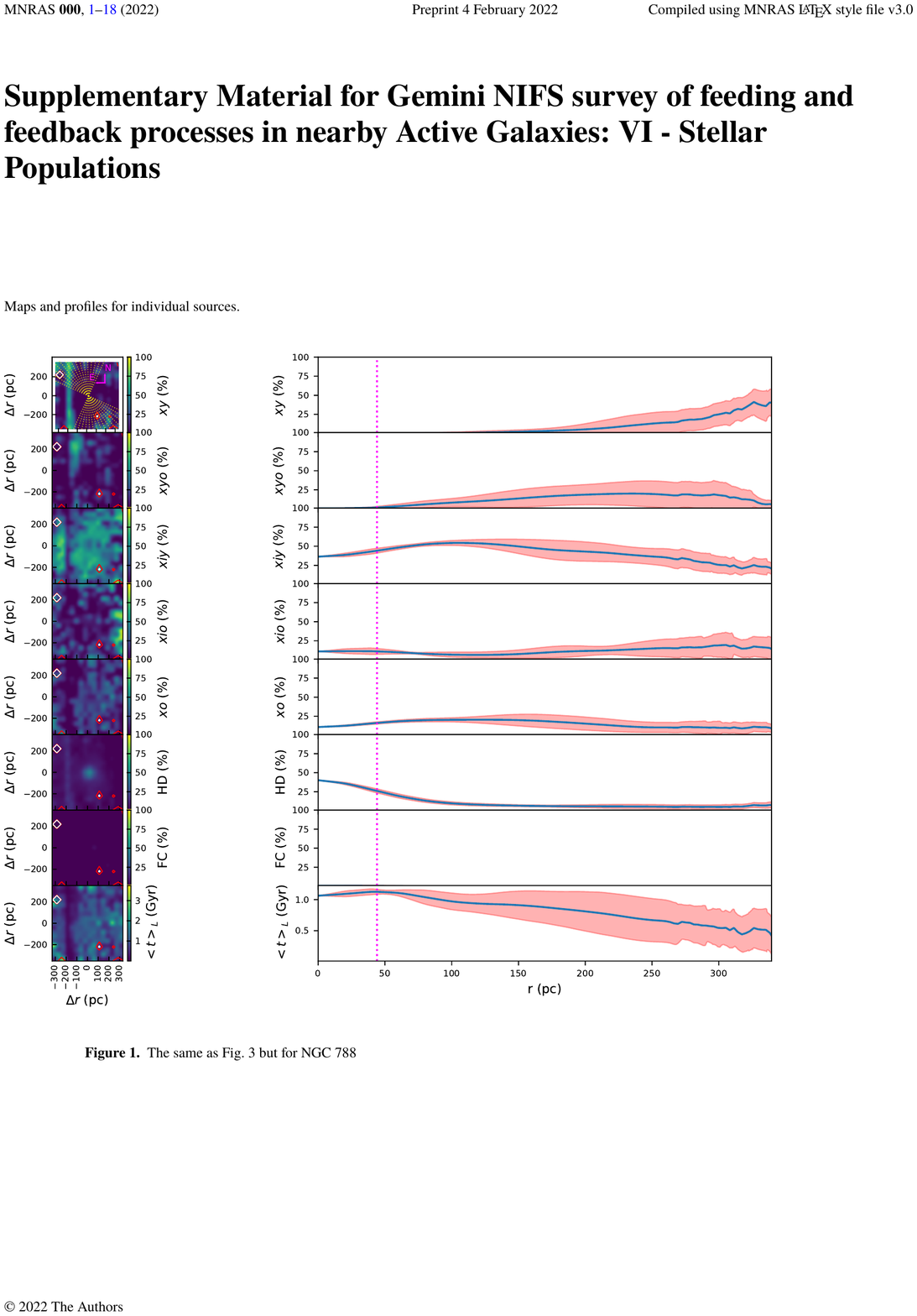}
\end{document}